\documentclass[prc,superscriptaddress,unsortedaddress,twocolumn,showpacs,preprintnumbers,amsmath,amssymb,floatfix]{revtex4}
\usepackage[dvips]{graphicx}
\usepackage{amsmath}
\usepackage{amssymb}
\usepackage{times}
\usepackage{mathrsfs}
\usepackage{multirow}
\usepackage{bm}
\usepackage{wrapft}
\usepackage{color}

\def\la{\mathrel{\mathpalette\fun <}}
\def\ga{\mathrel{\mathpalette\fun >}}
\def\fun#1#2{\lower3.6pt\vbox{\baselineskip0pt\lineskip.9pt
\ialign{$\mathsurround=0pt#1\hfil##\hfil$\crcr#2\crcr\sim\crcr}}}

\def\b{\beta}

\newcommand{\bequ}{\begin{equation}}
\newcommand{\eequ}{\end{equation}}
\newcommand{\bea}{\begin{eqnarray}}
\newcommand{\eea}{\end{eqnarray}}

\newcommand{\bfi}[1]{\mbox{\boldmath $#1$}}

\newcommand{\vb}{{\bfi b}}
\newcommand{\vk}{{\bfi k}}

\newcommand{\vrr}{{\bfi r}}
\newcommand{\vR}{{\bfi R}}

\begin{document}


\title{Effective radii of deuteron induced reactions}

\author{Shintaro Hashimoto}
\email[]{hashimoto.shintaro@jaea.go.jp}
\affiliation{
Advanced Science Research Center, Japan Atomic Energy Agency,
\\Ibaraki 319-1195, Japan}

\author{Masanobu Yahiro}
\affiliation{Department of Physics, Kyushu University, Fukuoka 812-8581, Japan}

\author{Kazuyuki Ogata}
\affiliation{Department of Physics, Kyushu University, Fukuoka 812-8581, Japan}

\author{Kosho Minomo}
\affiliation{Department of Physics, Kyushu University, Fukuoka 812-8581, Japan}

\author{Satoshi Chiba}
\affiliation{
Advanced Science Research Center, Japan Atomic Energy Agency,
\\Ibaraki 319-1195, Japan}

\date{\today}

\begin{abstract}
The continuum-discretized coupled-channels method (CDCC)
for exclusive reactions and
the eikonal reaction theory (ERT) as an extension of CDCC
to inclusive reactions are applied to deuteron induced reactions.
The CDCC result reproduces experimental data on the reaction cross section
for $d+^{58}$Ni scattering at 200~MeV/nucleon and ERT does data
on the neutron-stripping cross section for inclusive $^7$Li$(d,n)$ reaction
at 40~MeV. For deuteron induced reactions at 200~MeV/nucleon,
target-dependence of the reaction, elastic-breakup,
nucleon-stripping, nucleon-removal, complete- and incomplete-fusion
cross sections is clearly explained by simple formulae.
Accuracy of the Glauber model is also investigated.
\end{abstract}


\maketitle

\section{Introduction}

Understanding of the fusion reaction mechanism is one of the most
important and challenging subjects in nuclear physics.
The fusion reaction consists of complete and incomplete
fusion processes.
In the complete fusion process, all of the projectile is absorbed by
the target nucleus.
In the incomplete fusion process, meanwhile,
a part of the projectile is absorbed, while other part(s) of the projectile
is emitted.
The complete fusion process at low incident energies
is essential to understand the production of superheavy nuclei.
The incomplete fusion process in the scattering of unstable nuclei
at intermediate energies is important to extract information on
the projectile from the scattering.
Actually, the nucleon removal reaction widely used
for the spectroscopy of unstable nuclei~\cite{Gade} is composed of
the nucleon-stripping reaction
as a consequence of the incomplete fusion process and
the elastic-breakup reaction as a result of the direct reaction process.
Furthermore, the proton stripping process
in inclusive $^7$Li$(d,n)$ reaction  at incident energies
up to 50 MeV attracts
wide interests of not only nuclear physicists but also
nuclear engineers, because emitted neutrons through the process
are planned to be used
in the international fusion materials irradiation facility
(IFMIF)~\cite{Matsui}. Accurate evaluation of the proton-stripping cross
section is highly required.

The theoretical tool of analyzing the incomplete fusion process
at intermediate energies is the Glauber model~\cite{Glauber}.
The theoretical foundation of the model is investigated
in Ref.~\cite{Yahiro-Glauber}.
The Glauber model is based on the eikonal and the adiabatic
approximation; the latter is known to make
the elastic-breakup and removal cross sections diverge when
the Coulomb interaction is included; see for example Ref.~\cite{Capel-08}.
The Glauber model has thus been applied only for
lighter targets in which the Coulomb interaction is
negligible; see for example Refs.~\cite{Gade,Tostevin,Hencken,Ogawa01,Ibrahim,30Ne}.
Very recently, inclusive $^7$Li$(d,n)$ reaction at 40~MeV~\cite{Hagiwara}
was analyzed by the hybrid calculation~\cite{Ye}
in which the elastic-breakup component is evaluated by
the continuum-discretized coupled-channels method
(CDCC)~\cite{CDCC-review1,CDCC-review2} and the
proton stripping component is by the Glauber model.
The analysis was successful in reproducing the data~\cite{Hagiwara},
even if the Coulomb interaction is neglected in the Glauber-model
calculation.
Such a hybrid calculation should be justified
by more accurate reaction theories.

CDCC is an accurate method for treating exclusive reactions such as
elastic scattering and elastic-breakup reactions.
The theoretical foundation of CDCC is shown
in Refs.~\cite{CDCC-foundation1,CDCC-foundation2,CDCC-foundation3}.
CDCC succeeded in reproducing data on the scattering of
stable and unstable projectiles~\cite{CDCC-review1,CDCC-review2,
Rusek,Tostevin2,Davids,Mortimer,Eikonal-CDCC,Matsumoto3,Howell,Rusek2,
Matsumoto4,Moro,THO-CDCC,4body-CDCC-bin,Matsumoto:2010mi,Avrigeanu}.
Very recently, CDCC was extended to inclusive reactions such as
nucleon stripping reactions~\cite{ERT}. This method is referred to as
the eikonal reaction theory (ERT).
In ERT, the adiabatic approximation is not made for the Coulomb interaction,
so that the elastic-breakup and the nucleon removal reaction never diverge.
ERT is thus applicable for both light and heavy targets.
ERT assumes the eikonal approximation to be good.
The formulation starts with the eikonal approximation, but
non-eikonal corrections are made by calculating
fusion cross sections with CDCC.
This is essential progress in the theory on fusion reactions,

Extensive measurements of total-reaction and
nucleon-removal cross sections are now being made
for the scattering of unstable nuclei
at intermediate energies, say 100-300~MeV/nucleon, in MSU, RIKEN, and GSI.
Accurate understanding of the fusion reaction mechanism is thus
highly required at intermediate energies.
In this paper, we mainly consider deuteron induced reactions
at 200~MeV/nucleon as a typical case and
analyze integrated cross sections of the reactions with CDCC and ERT.
Deuteron is fragile just as unstable nuclei and furthermore
it has no ambiguity on the structure. In this sense, deuteron is the
most suitable projectile to understand the fusion reaction mechanism.
We will show that CDCC reproduces experimental data
on the reaction cross section
for $d+^{58}$Ni scattering at 200~MeV/nucleon and ERT does data
on the neutron-stripping cross section for inclusive $^7$Li$(d,n)$ reaction
at 40~MeV. Target-mass-number ($A$) dependence of the
reaction, elastic-breakup,
nucleon-removal, nucleon-stripping, incomplete- and complete-fusion
cross sections for deuteron induced reactions at 200~MeV/nucleon
is clearly explained with simple formulae.
Accuracy of the Glauber model will be investigated.

ERT is recapitulated in Sec.~\ref{Eikonal reaction theory}.
Numerical results of CDCC and ERT are presented
in Sec.~\ref{Numerical results}.
Section~\ref{Summary} is devoted to summary.

\section{Eikonal reaction theory}
\label{Eikonal reaction theory}

\subsection{Three-body model}
\label{Three-body model}

Deuteron ($d$) is the system in which proton ($p$) and neutron ($n$) are
weakly bound. It is thus natural to assume that
scattering of $d$ from target T is well described by
the $p+n+$T three-body model.
Actually, the model is successful in reproducing the
experimental data on elastic scattering and
breakup reactions of $d$~\cite{CDCC-review1,CDCC-review2}.
The model Hamiltonian is
\bea
  H= -\frac{\hbar^2}{2\mu}\nabla_R^2+ U(r_p,r_n) + h
\label{Model-Hamiltonian}
\eea
with
\bea
 U(r_p,r_n) = U_p^{\rm (N)}(r_p)
            + U_p^{\rm (C)}(r_p) + U_n^{\rm (N)}(r_n) ,
\label{Pot}
\eea
where $h=T_r +V(\vrr)$ denotes the intrinsic Hamiltonian of $d$ that
consists of the kinetic-energy operator $T_r$ and the interaction $V$.
Furthermore, $\mu$ is the reduced mass between $d$ and T,
$U_p^{\rm (N)}$ ($U_n^{\rm (N)}$) is the nuclear part of
the proton (neutron) optical potential and $U_p^{\rm (C)}$ is
the Coulomb interaction between $p$ and T.
The three-dimensional vector $\vR=(\vb,Z)$ stands
for the coordinate between $d$  and T, while
$\vrr$ is the coordinate between $p$ and $n$.
The vector $\vrr_x=(\vb_x,z_x)$ for $x=p$ or $n$
is the coordinate between $x$ and T.
The total wave function $\Psi(\vR,\vrr)$ of the three-body system is
then obtained by solving the three-body Schr\"odinger equation
\bea
   [ H-E]\Psi(\vR,\vrr)=0 .
\label{Schrodinger-eq}
\eea

In the three-body model, transitions of the incident flux to
the inelastic (target-excitation) channels are expressed
by the imaginary parts of $U_p^{\rm (N)}$
and $U_n^{\rm (N)}$.
The imaginary part of
$U_p^{\rm (N)}$ denotes the absorption of $p$ by T,
while the imaginary part of $U_n^{\rm (N)}$ corresponds
to the absorption of $n$ by T.
Therefore, the three-body model implicitly assumes that
$p$ and $n$ are absorbed independently.

\subsection{Separation of $S$-matrix}
\label{Separation of S-matrix}

We consider $d$ scattering at intermediate energies,
say 200~MeV/nucleon. Since the eikonal approximation is considered to be
good for the scattering, the $S$-matrix elements and several types of
cross sections are described by ERT. Accuracy of the eikonal approximation
is investigated later.
In this subsection we recapitulate ERT for deuteron scattering.
In the eikonal approximation, the three-body wave function $\Psi$
is assumed to be
\bea
    \Psi={\hat O} \psi(\vR,\vrr)
\label{product}
\eea
with the operator
\bea
{\hat O} =\frac{1}{\sqrt{\hbar {\hat v}}} e^{i {\hat K} \cdot Z} ,
\eea
where ${\hat K}=\sqrt{2\mu(E-h)}/{\hbar}$
and ${\hat v}={\hbar {\hat K}}/{\mu}$  are wave-number and velocity operators
of the motion of deuteron relative to T, respectively.
When Eq.~\eqref{product} is inserted into Eq.~\eqref{Schrodinger-eq},
we have a term including $\nabla_R^2\psi$ but it is neglected,
since $\psi$ is slowly varying with $\vR$ compared
with ${\hat O}$. The neglect leads Eq.~\eqref{Schrodinger-eq} to
\bea
  i \frac{d\psi}{dZ}={\hat O^{\dagger}}U{\hat O}\psi .
\label{eikonal-eq}
\eea
Regarding $Z$ as ``time" and solving Eq.~\eqref{eikonal-eq}
iteratively, we obtain the formal solution
\bea
\psi=\exp\Big[
-i{\cal P}\int_{-\infty}^Z dZ' {\hat O^{\dagger}}U{\hat O}
\Big] ,
\label{WF}
\eea
where ${\cal P}$ is the ``time" ordering operator. Taking $Z$ to $\infty$
in Eq.~\eqref{WF}, we get the $S$-matrix operator
\bea
S=\exp\Big[
-i{\cal P}\int_{-\infty}^{\infty} dZ {\hat O^{\dagger}}U{\hat O}
\Big]  .
\label{S-matrix-operator}
\eea

In the Glauber model,
$h$ is replaced by the ground-state energy $\epsilon_0$ of $d$.
This adiabatic approximation reduces ${\hat O^{\dagger}}U{\hat O}$ and
${\cal P}$ in Eq.~\eqref{S-matrix-operator}
to $U/(\hbar v_0)$ and 1, respectively,
where $v_0$ is the velocity of $d$
in the ground state relative to T.
This is nothing but the $S$-matrix in the Glauber model.
Thus, ERT is an extension of the Glauber model.

The operator ${\hat O^{\dagger}}U{\hat O}$ describes the change in the
motion of
$p$ and $n$ in $d$ during the collision.
The change is small for the short-range nuclear interactions,
$U_p^{\rm (N)}$ and $U_n^{\rm (N)}$, while large for
the long-range Coulomb interaction $U_p^{\rm (C)}$.
Therefore, the adiabatic approximation that neglects this change
is good for the nuclear interactions but not for
the Coulomb interaction.

More quantitative discussion can be made by considering
the matrix element
\bequ
\int_{-\infty}^{\infty} dZ
\langle \phi_{\vk} | {\hat O^{\dagger}}U{\hat O} | \phi_{0} \rangle
\approx
\frac{ e^{i (K_0 - K) R_U} }{\hbar v_0}
\int_{-\infty}^{\infty} dZ
\langle \phi_{\vk} | U | \phi_{0} \rangle
\label{Estimation}
\eequ
between the ground state $\phi_{0}$ of $d$ with the intrinsic
energy $\epsilon_0$ and its continuum state $\phi_{\vk}$
with the intrinsic momentum $\hbar \vk$ and energy $\epsilon$,
where $\hbar K_0$ ($\hbar K$)
is the momentum of $d$ in the ground (continuum) state relative to T,
and $R_U$ is the range of the interaction considered.
As an example, let us consider the $d+^{208}$Pb scattering at 200~MeV/nucleon.
The spectrum of the elastic breakup reaction, $d\sigma_{\rm EB}/d\epsilon$,
has a peak around $\epsilon=10$~MeV. The interaction range
$R_U$ is about 7.1~fm for the nuclear interactions,
$U_p^{\rm (N)}$ and $U_n^{\rm (N)}$, while
that for the Coulomb interaction $U_p^{\rm (C)}$ is infinity.
Hence,
$\Delta=(K_0 - K) R_U \approx 0.55 < \pi$ for the nuclear interactions,
but $\infty$ for $U_p^{\rm (C)}$.
Since the adiabatic approximation is good for $\Delta \ll \pi$,
the approximation is acceptable for the nuclear interactions,
but not for the Coulomb interaction.
Actually, the breakup cross section is known to diverge
in the adiabatic approximation~\cite{Eikonal-CDCC,Capel-08}.

The fact that the adiabatic approximation is fairly good for
$U_n^{\rm (N)}$ indicates that
$U_n^{\rm (N)}$ is commutable with
${\hat O}$. Therefore, we can make the replacement
\bea
    {\hat O^{\dagger}}U_n^{\rm (N)} {\hat O}
    \leftrightarrow U_n^{\rm (N)}/(\hbar v_0) .
\label{replacement-n}
\eea
The accuracy of Eq.~\eqref{replacement-n} is confirmed later by
numerical calculations. Using Eq.~\eqref{replacement-n}, we get
\bea
  S=S_nS_p
  \label{S-separation}
\eea
with
\bea
  S_n&=&\exp\Big[
  -i {\cal P} \int_{-\infty}^{\infty} dZ O^{\dagger} U_n^{\rm (N)}
  {\hat O}  \Big] ,
  \label{Sn} \\
  S_p&=&\exp\Big[
-i{\cal P}\int_{-\infty}^{\infty} dZ {\hat O^{\dagger}}
(U_p^{\rm (N)}+U_p^{\rm (C)}) {\hat O} \Big] .
\label{Sp}
\eea
Thus, $S$ can be separated into the neutron and proton parts,
$S_n$ and $S_p$, respectively. The neutron part $S_n$
describes scattering of $n$ by $U_n^{\rm (N)}$ and recoil of $p$ by
the scattering. However, a velocity caused by the recoil
is much smaller than the initial velocity $v_0$ of $p$, so that the recoil
effect is negligible.
Similar interpretation is possible for $S_p$.
The operator $S_p$ is the formal solution of the Schr\"odinger equation
\bea
   \left[ -\frac{\hbar^2}{2\mu}\nabla_R^2 + h + U_p^{\rm (N)}(r_p)
   + U_p^{\rm (C)}(r_p)-E\right]\Psi_p=0 .
\label{Schrodinger-eq-p}
\eea
and $S_n$ is the solution of the Schr\"odinger equation
\bea
   \left[ -\frac{\hbar^2}{2\mu}\nabla_R^2 + h + U_n^{\rm (N)}(r_n)
   -E\right]\Psi_n=0 .
\label{Schrodinger-eq-n}
\eea
Hence, the matrix elements of $S_n$ and $S_p$ can be obtained
by solving Eqs.~\eqref{Schrodinger-eq-p} and \eqref{Schrodinger-eq-n}
with eikonal-CDCC~\cite{Eikonal-CDCC} in which
the eikonal approximation is made in CDCC calculations.
Non-eikonal corrections to $S_n$ and $S_p$ can be easily made
by using CDCC instead of eikonal-CDCC
in solving Eqs.~\eqref{Schrodinger-eq-p} and \eqref{Schrodinger-eq-n}.

\subsection{Integrated cross sections}
\label{Integrated cross sections}

In this subsection, we derive several formulae
of integrated cross sections
with the product form \eqref{S-separation},
following the formulation of the cross sections in
the Glauber model~\cite{Hussein,Hencken}.
The reaction and elastic-breakup cross sections,
$\sigma_{\rm R}$ and $\sigma_{\rm EB}$, respectively, are defined by
\bea
\sigma_{\rm R}&=&
\int d^2 \vb [1-
| \langle \phi_{0} | S_pS_n | \phi_{0} \rangle|^2] ,
\label{reaction-Xsec} \\
\sigma_{\rm EB}&=&
\int d^2 \vb [
\langle \phi_{0} | |S_pS_n|^2 | \phi_{0} \rangle -
|\langle \phi_{0} | S_pS_n | \phi_{0} \rangle|^2] .~~~
\label{breakup-Xsec}
\eea
The cross sections $\sigma_{\rm R}$ and $\sigma_{\rm EB}$ can
be evaluated from the asymptotic form of $\Psi$ that
is obtained by solving Eq.~\eqref{Schrodinger-eq} with CDCC.

The total fusion cross section $\sigma_{\rm TF}$ is
defined by
\bea
\sigma_{\rm TF}=\sigma_{\rm R}-\sigma_{\rm EB}
= \int d^2 \vb
\langle \phi_{0} |[(1-|S_pS_n|^2)]| \phi_{0} \rangle .
   \label{TF-Xsec}
\eea
The total fusion cross section can be decomposed into
the neutron-stripping cross section $\sigma_{n:{\rm STR}}$,
the proton-stripping cross section $\sigma_{p:{\rm STR}}$
and the complete-fusion cross section $\sigma_{\rm CF}$:
\bea
\sigma_{\rm TF}= \sigma_{n:{\rm STR}}+\sigma_{p:{\rm STR}}+\sigma_{\rm CF},
   \label{TF-Xsec-decom}
\eea
where
\bea
\sigma_{n:{\rm STR}}
&=&\int d^2 \vb
\langle \phi_{0} | |S_p|^2(1-|S_n|^2) | \phi_{0} \rangle ,
\label{n:STR-Xsec} \\
\sigma_{p:{\rm STR}}
&=&\int d^2 \vb
\langle \phi_{0} | |S_n|^2(1-|S_p|^2) | \phi_{0} \rangle ,
\\
\sigma_{\rm CF}
&=&\int d^2 \vb
\langle \phi_{0} |(1-|S_n|^2)(1-|S_ p|^2) | \phi_{0} \rangle .
\eea
The factor $|S_p|^2(1-|S_n|^2)$ in $\sigma_{n:{\rm STR}}$
shows that $p$ is scattered by T while $n$ is absorbed by T, and
the factor $(1-|S_n|^2)(1-|S_p|^2)$
in $\sigma_{\rm CF}$ means that both $p$ and $n$ are absorbed by T.
The sum of $\sigma_{n:{\rm STR}}$ and $\sigma_{p:{\rm STR}}$ describes
the incomplete fusion cross section $\sigma_{\rm IF}$:
\bea
\sigma_{\rm IF}= \sigma_{n:{\rm STR}}+\sigma_{p:{\rm STR}}.
   \label{TF-Xsec-decom-2}
\eea

In the neutron removal reaction,
$n$ is either absorbed or scattered by T,
while $p$ is independently scattered by T.
Hence, the cross section $\sigma_{-n}$ is the sum of
$\sigma_{\rm EB}$ and $\sigma_{n:{\rm STR}}$:
\bea
\sigma_{-n}=\sigma_{\rm EB}+\sigma_{n:{\rm STR}} .
\eea
The neutron-stripping cross section $\sigma_{n:{\rm STR}}$
is rewritten into
\bea
\sigma_{n:{\rm STR}}
&=&\int d^2 \vb
\langle \phi_{0} |[(1-|S_pS_n|^2) -(1-|S_p|^2)]
| \phi_{0} \rangle
\nonumber \\
&=& \sigma_{\rm TF}-\sigma_{\rm TF}(p) ,
\label{neutron-removal-Xsec}
\eea
where
\bea
\sigma_{\rm TF}(p)&=&\sigma_{\rm R}(p)-\sigma_{\rm EB}(p)
\eea
with
\bea
\sigma_{\rm R}(p)&=&
\int d^2 \vb [1-
| \langle \phi_{0} | S_p | \phi_{0} \rangle|^2] ,
\label{Sigma-R-p} \\
\sigma_{\rm EB}(p)&=&
\int d^2 \vb [
\langle \phi_{0} | |S_p|^2 | \phi_{0} \rangle -
|\langle \phi_{0} | S_p| \phi_{0} \rangle|^2] .~~~
\label{Sigma-Bu-p}
\eea
Here, $\sigma_{\rm TF}(p)$, $\sigma_{\rm R}(p)$ and $\sigma_{\rm EB}(p)$
are the total fusion, reaction and elastic-breakup cross sections
induced by $U_p^{\rm (N)}+U_p^{\rm (C)}$ only.
The cross sections, $\sigma_{\rm R}(p)$ and $\sigma_{\rm EB}(p)$,
can be evaluated from the asymptotic form of $\Psi_p$ that
are obtained by solving Eq.~\eqref{Schrodinger-eq-p} with CDCC.
Thus, we can evaluate $\sigma_{n:{\rm STR}}$ with Eq.~\eqref{neutron-removal-Xsec}.

Similarly, the proton removal cross section $\sigma_{-p}$ is
obtained by
\bea
\sigma_{-p}=\sigma_{\rm EB}+\sigma_{p:{\rm STR}}
\eea
and the proton-stripping cross section $\sigma_{p:{\rm STR}}$ is
rewritten into
\bea
\sigma_{p:{\rm STR}}
&=& \sigma_{\rm TF} - \sigma_{\rm TF}(n),
\label{proton-removal-Xsec}
\eea
where
\bea
\sigma_{\rm TF}(n)&=&\sigma_{\rm R}(n)-\sigma_{\rm EB}(n)
\eea
with
\bea
\sigma_{\rm R}(n)&=&
\int d^2 \vb [1-
| \langle \phi_{0} | S_n | \phi_{0} \rangle|^2] ,
\label{Sigma-R-n} \\
\sigma_{\rm EB}(n)&=&
\int d^2 \vb [
\langle \phi_{0} | |S_n|^2 | \phi_{0} \rangle -
|\langle \phi_{0} | S_n| \phi_{0} \rangle|^2] .~~~
\label{Sigma-EB-n}
\eea
The cross sections, $\sigma_{\rm R}(n)$ and $\sigma_{\rm EB}(n)$,
can be evaluated from the asymptotic form of $\Psi_n$ that
are obtained by solving Eq.~\eqref{Schrodinger-eq-n}
with CDCC.
We can thus evaluate $\sigma_{p:{\rm STR}}$ with Eq.~\eqref{proton-removal-Xsec}.
Finally, $\sigma_{\rm CF}$ is obtained from
$\sigma_{\rm TF}$ and $\sigma_{\rm IF}=\sigma_{n:{\rm STR}}+\sigma_{p:{\rm STR}}$
by using the relation $\sigma_{\rm CF}=\sigma_{\rm TF}-\sigma_{\rm IF}$.

\subsection{Tests of the eikonal and the adiabatic approximation}
\label{test of approximations}

In ERT, non-eikonal corrections to the integrated cross sections are
taken into account by using CDCC instead of eikonal-CDCC.
For the $d+^9$Be scattering at 200~MeV/nucleon,
the correction is found to be less than 1\% for $\sigma_{\rm R}$,
$\sigma_{\rm EB}$, $\sigma_{n:{\rm STR}}$ and $\sigma_{p:{\rm STR}}$.
For the $d+^{208}$Pb scattering at 200~MeV/nucleon,
the correction is 1.5\% for $\sigma_{\rm R}$,
$\sigma_{\rm EB}$ and $\sigma_{n:{\rm STR}}$ and 16\% for $\sigma_{p:{\rm STR}}$.
Thus, the non-eikonal correction is small except for
$\sigma_{p:{\rm STR}}$ for heavy targets.
As shown in Eq.~\eqref{proton-removal-Xsec}, $\sigma_{p:{\rm STR}}$ is
approximately obtained
by
\bea
\sigma_{p:{\rm STR}} \approx \sigma_{\rm R}-\sigma_{\rm R}(n) ,
\eea
since $\sigma_{\rm EB} \ll \sigma_{\rm R}$.
The 1.5\% correction appears in $\sigma_{\rm R}$ because of
the strong Coulomb field, while the correction is negligible
in $\sigma_{\rm R}(n)$ as a consequence of the absence of the Coulomb field.
Thus, one can conclude that the 16\% correction required for
$\sigma_{p:{\rm STR}}$ is nothing but the
1.5\% correction for $\sigma_{\rm R}$.
Note that $\sigma_{p:{\rm STR}}$ is much smaller than $\sigma_{\rm R}$.
Meanwhile, $\sigma_{n:{\rm STR}}$ is given by
\bea
\sigma_{n:{\rm STR}} \approx \sigma_{\rm R}-\sigma_{\rm R}(p)
\eea
The 1.5~\% corrections appear in both $\sigma_{\rm R}$ and
$\sigma_{\rm R}(p)$. The cancellation between
the two corrections makes the non-eikonal correction small for
$\sigma_{n:{\rm STR}}$.

In ERT, the adiabatic approximation is assumed to be good for
the nuclear potential $U_n^{\rm (N)}$.
This can be tested by setting
\[
U(R,r_n)=U_p^{\rm (C)}(R)+U_p^{\rm (N)}(R)+U_n^{\rm (N)}(r_n)
\]
in the Schr\"odinger equation \eqref{Schrodinger-eq}.
In this setup,
the projectile breakup is induced only by $U_n^{\rm (N)}(r_n)$, since
the argument $r_p$ of $U_p^{\rm (C)}$ and $U_p^{\rm (N)}$
has been replaced by $R$.
Switching the adiabatic approximation on the Schr\"odinger equation
corresponds to the replacement \eqref{replacement-n}.
For the $d+^{9}$Be scattering at 200~MeV/nucleon,
the error due to the approximation
is 0.3\% for $\sigma_{\rm R}$ and 2\% for $\sigma_{\rm EB}$.
For the $d+^{208}$Pb scattering at 200~MeV/nucleon,
the error is 0.4\% for $\sigma_{\rm R}$ and 6\% for
$\sigma_{\rm EB}$.
Errors due to these approximations are even smaller for heavier projectiles
such as $^{31}$Ne~\cite{ERT}.

\section{Numerical results}
\label{Numerical results}

We use the Koning-Delaroche global optical potential~\cite{KDpot}
as $U_p^{\rm (N)}$ and $U_n^{\rm (N)}$, and
the Ohmura potential~\cite{Ohmura} as $V$ that
reproduces the deuteron binding energy $|\varepsilon_0|=2.23$ MeV.
As the model space of CDCC calculation,
s-, p-, and d-wave breakup states with $k \le 1.0$~${\rm fm}^{-1}$ are taken.
Each $k$-continuum is divided into small bins
with a common width $\Delta k =0.1$~fm$^{-1}$, and
the breakup states within each bin are averaged into a single state.
Maximum values of $r$ and $R$ are $r_{\rm max}=$ 200~fm and
$R_{\rm max}=$ 200~fm, respectively.

\subsection{$d+^{58}$Ni elastic scattering at 400~MeV}
\label{sec:Elastic scattering}

In this subsection,
we consider the $d+^{58}$Ni elastic scattering at
200~MeV/nucleon, because the elastic cross section was measured and
the reaction cross section was evaluated
with the optical potential analysis~\cite{d58Ni}.

\begin{figure}[htbp]
\begin{center}
 \includegraphics[width=0.4\textwidth,clip]{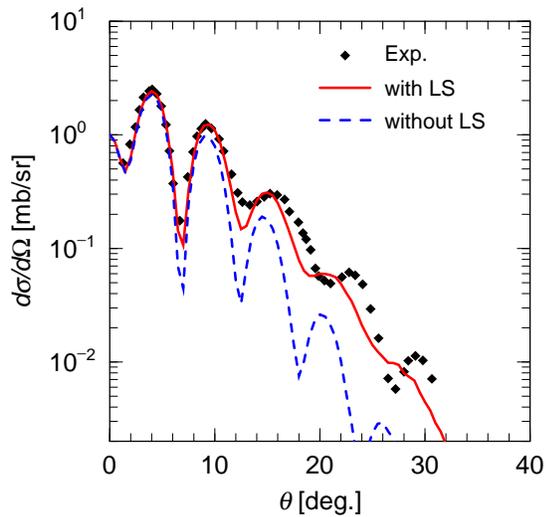}
 \caption{(Color online)
 The elastic cross section of $d+^{58}$Ni scattering at 200~MeV/nucleon
 as a function of the c.m. scattering angle $\theta$.
 The solid (dashed) line stands for result of CDCC calculation with
 (without) the spin-orbit interactions of the proton and neutron
 optical potentials.
 The experimental data are taken from Ref.~\cite{d58Ni}.
  }
 \label{Fig-Spin-orbit}
\end{center}
\end{figure}

Figure \ref{Fig-Spin-orbit} shows the elastic cross section
as a function of the center-of-mass (c.m.) scattering angle $\theta$.
The solid (dashed) line represents a result of CDCC calculation
with (without) the spin-orbit interactions of $U_p^{\rm (N)}$ and
$U_n^{\rm (N)}$.
The solid line well reproduces the experimental data.
Large deviation of the dashed line from the solid line
for $\theta \ga 10^\circ$
shows that
the spin-orbit interactions yield a significant effect on the
elastic cross section.

\begin{figure}[htbp]
\begin{center}
 \includegraphics[width=0.4\textwidth,clip]{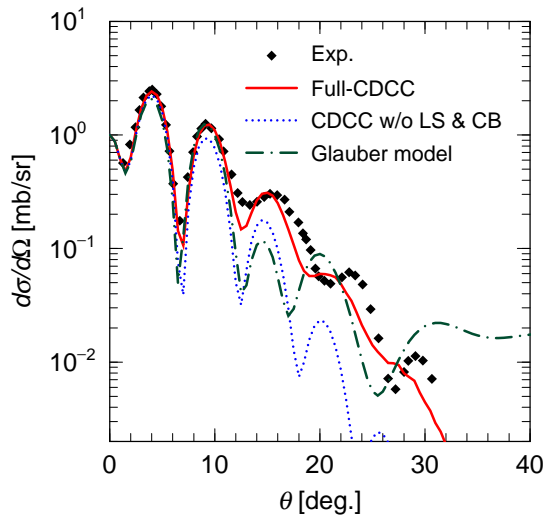}
 \caption{(Color online)
 Comparison of the Glauber model with CDCC for the elastic cross section
 of $d+^{58}$Ni scattering at 200~MeV/nucleon.
 The solid line is the same as the solid line in Fig.~\ref{Fig-Spin-orbit}.
 The dotted line stands for the result of CDCC calculation
 with neglecting both spin-orbit interactions and Coulomb breakup.
 The dot-dashed line represents the result of the Glauber-model calculation.
 The experimental data are taken from Ref.~\cite{d58Ni}.
  }
 \label{Fig-Glauber}
\end{center}
\end{figure}

In the Glauber model, the spin-orbit interactions and
the Coulomb breakup are neglected, and furthermore
the eikonal and the adiabatic approximation are made.
The Coulomb breakup can be neglected by replacing
$U_p^{\rm (C)}(r_p)$ by $U_p^{\rm (C)}(R)$
in CDCC calculation. In Fig.~\ref{Fig-Glauber},
the dotted line is the result of CDCC calculation
with neglecting both the spin-orbit interactions and the Coulomb breakup.
The dotted line agrees with the dashed line of Fig.~\ref{Fig-Spin-orbit},
that is, the result of
CDCC calculation with the Coulomb breakup and without spin-orbit interactions.
Thus, the Coulomb breakup effect is small.
In Fig.~\ref{Fig-Glauber}, the dot-dashed line represents a result of the
Glauber-model calculation. The large deviation of the dot-dashed line
from the dotted line comes from the eikonal and adiabatic approximations,
more precisely from the eikonal approximation.
Eventually, the result of the Glauber model (the dot-dashed line)
largely deviates from the full-CDCC result (the solid line) in which
both the spin-orbit interactions and the Coulomb breakup are
taken into account.
Thus, the Glauber model does not work well for the elastic
cross section for $\theta \ga 10^\circ$.

The reaction cross section calculated by full-CDCC is 1056~mb, while
the value extracted from
the measured elastic cross section with an optical model analysis
is 1083~mb~\cite{d58Ni}.
Thus, the CDCC result is consistent with the experimental data.
Table~\ref{tbl-d58Ni} shows effects of
the spin-orbit interactions and the Coulomb breakup on
$\sigma_{\rm R}$, $\sigma_{\rm EB}$ and $\sigma_{\rm TF}$.
Comparing the results with each other,
one can find that
the Coulomb breakup effect is about 50\% for $\sigma_{\rm EB}$ and
2\% for $\sigma_{\rm R}$, while the spin-orbit interaction effect is
17\% for $\sigma_{\rm EB}$ and 1\% for $\sigma_{\rm R}$.
Thus, the effects are sizable for $\sigma_{\rm EB}$ and
appreciable for $\sigma_{\rm R}$.
Meanwhile, these effects are quite small for $\sigma_{\rm TF}$,
because the absolute value of $S$ is mainly determined by
the imaginary parts of $U_n^{\rm (N)}$ and $U_p^{\rm (N)}$.
This is also the case with
$\sigma_{p:{\rm STR}}$, $\sigma_{n:{\rm STR}}$,
$\sigma_{\rm IF}$ and $\sigma_{\rm CF}$.
Table~\ref{tbl-d58Ni} shows that the spin-orbit interaction effect
is even smaller than the Coulomb breakup effect.
We henceforth neglect the spin-orbit interactions
but not the Coulomb breakup, since the Coulomb breakup
becomes more significant for heavier targets.

\begin{table}[htbp]
\begin{center}
\begin{tabular}{c|cc|ccc}
 \hline \hline
 & Coulomb breakup & spin-orbit &
 $\sigma_{\rm R}$ & $\sigma_{\rm EB}$ & $\sigma_{\rm TF}$ \\
 \hline
 CDCC & on  & on  & 1056 & 59 & 997 \\ \cline{2-6}
      & on  & off & 1066 & 69 & 997 \\ \cline{2-6}
      & off & on  & 1029 & 31 & 998 \\ \cline{2-6}
      & off & off & 1023 & 25 & 998 \\ \cline{2-6}
 \hline
 Exp. & & & 1083 & & \\
 \hline \hline
\end{tabular}
\end{center}
\caption{\label{tbl-d58Ni}
Effects of the spin-orbit interaction and the Coulomb breakup
on $\sigma_{\rm R}$, $\sigma_{\rm EB}$ and $\sigma_{\rm TF}$
for the $d+^{58}$Ni scattering at 200~MeV/nucleon.
The experimental data is taken from Ref.~\cite{d58Ni}.
The cross sections are shown in units of mb.
}
\end{table}

\subsection{Inclusive $^{7}$Li$(d,n)$ reaction at 40~MeV}
\label{sec:d+Li}

The double differential cross section (DDX) of inclusive $^{7}$Li$(d,n)$
reaction was measured
at 40~MeV~\cite{Hagiwara}.
The main part of the DDX consists of the elastic-breakup and the proton
stripping parts. When the elastic-breakup DDX is calculated with CDCC
and subtracted from the measured DDX,
the angular and the energy dependence of the
remaining DDX is well reproduced by the Glauber model~\cite{Ye}.
This indicates that the proton-stripping cross section
$\sigma_{p:{\rm  STR}}$ can be obtained by fitting the theoretical DDX
calculated by the Glauber model to the remaining DDX, and integrating
it over the angle and the energy. The $\sigma_{p:{\rm STR}}$ thus extracted is
$244 \pm 34({\rm theor.}) \pm 37({\rm exp.})$~mb;
the theoretical error comes from ambiguity of the fitting.
ERT gives $\sigma_{p:{\rm  STR}}=253$~mb and the Glauber model
does 214~mb. Thus, the two theoretical results are consistent with
the experimental data. The ERT result seems to be slightly better than
the Glauber-model result for this case.

\subsection{Relation between reaction and elastic-breakup cross sections}
\label{sec:relation-R-EB}

In this subsection, we discuss the relation between
$\sigma_{\rm R}$ and $\sigma_{\rm EB}$
for deuteron induced reactions at 200~MeV/nucleon.

In the framework of CDCC,
$\sigma_{\rm R}$ is the sum of
the partial reaction cross section $\sigma_{\rm R}(J)$ over the total
angular momentum $J$, while
$\sigma_{\rm EB}$ is the sum of
the partial breakup cross sections $\sigma_{\rm EB}(J)$:
\bea
\sigma_{\rm R}&=&\sum_{J} \sigma_{\rm R}(J)
=\frac{\pi}{K_0^2} \sum_{J} (2J+1)  P_{\rm R}(J) ,
\label{Xsec-R}
\\
\sigma_{\rm EB}&=& \sum_{J} \sigma_{\rm EB}(J)
=\frac{\pi}{K_0^2} \sum_{J} (2J+1) P_{\rm EB}(J)
\label{Xsec-EB}
\eea
with
\bea
P_{\rm R}(J) &=& 1-|\langle 0 |S(J) | 0 \rangle |^2  ,
\\
P_{\rm EB}(J) &=& \sum_{\b} |\langle \b |S(J) | 0 \rangle |^2 ,
\eea
where $K_0$ is the initial wave number of $d$.
The partial elastic and breakup $S$-matrix elements
are denoted by $\langle 0 |S(J) | 0 \rangle$
and $\langle \b |S(J) | 0 \rangle$, respectively, where $0$ ($\b$)
represents the elastic (breakup) channel.
The quantity $P_{\rm R}(J)$ shows, for each $J$,
the transition probability of
the incident flux to all channels except the elastic channel,
while $P_{\rm EB}(J)$ describes the transition probability to all
the breakup channels.
The probability $P_{\rm EB}(J)$ can be rewritten into
\bea
P_{\rm EB}(J) = \langle 0 | S(J)^{\dagger}S(J) | 0 \rangle
              - |\langle 0 | S(J) | 0 \rangle|^2 .
\eea
This indicates that $P_{\rm EB}(J)$ is the fluctuation of
the mean value $|\langle 0 | S(J) | 0 \rangle|$ for each $J$.
In general, a rapid change in $|\langle 0 | S(J) | 0 \rangle|$
with respect to $J$ occurs where the fluctuation becomes maximum.
Since $P_{\rm R}(J)$ is a function of $|\langle 0 | S(J) | 0 \rangle|$,
one can expect that $P_{\rm R}(J)$ is rapidly changed
where $P_{\rm EB}(J)$ becomes maximum. We return to this point below.

\begin{figure}[htbp]
\begin{center}
 \includegraphics[width=0.4\textwidth,clip]{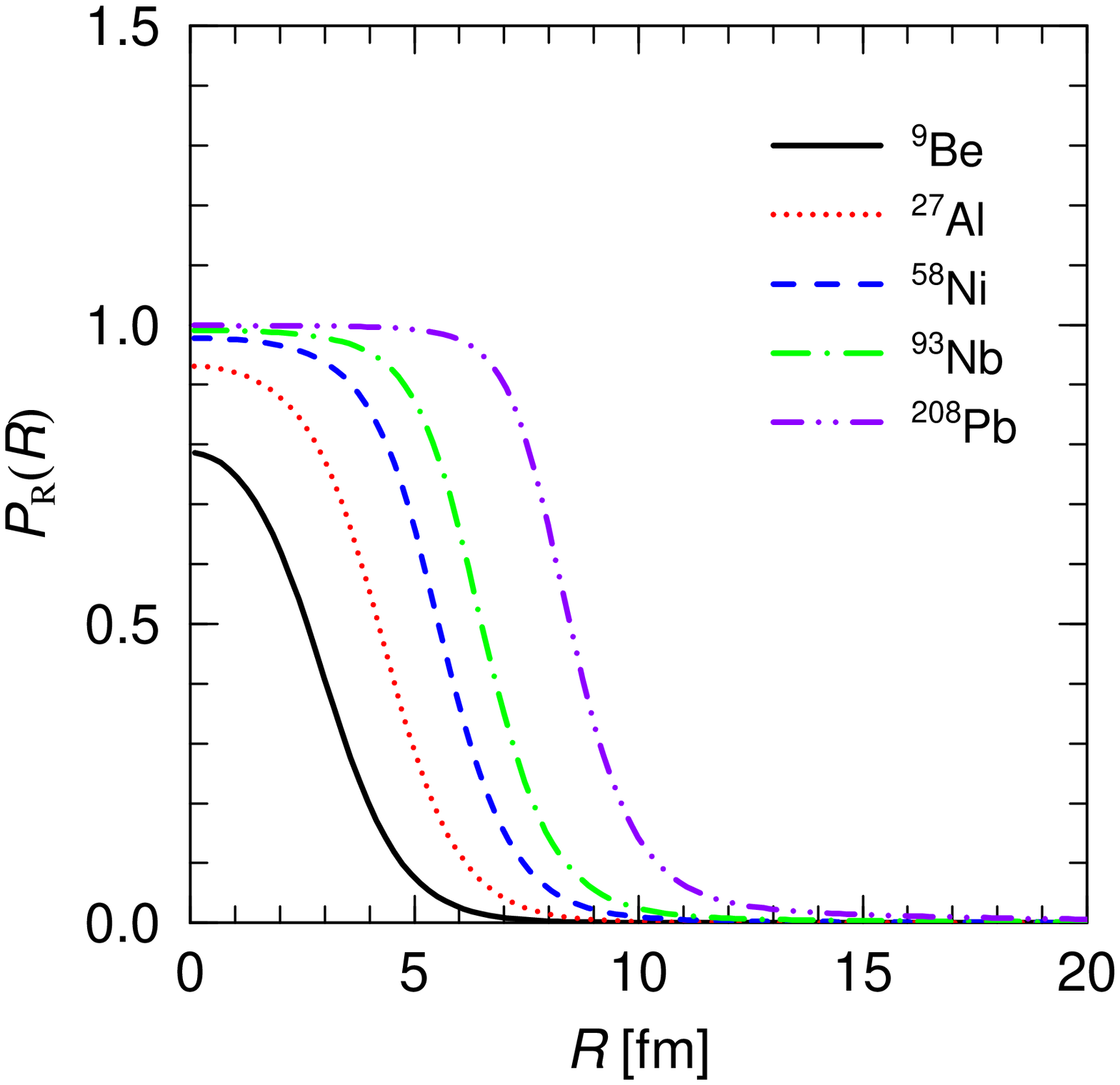}
 \includegraphics[width=0.4\textwidth,clip]{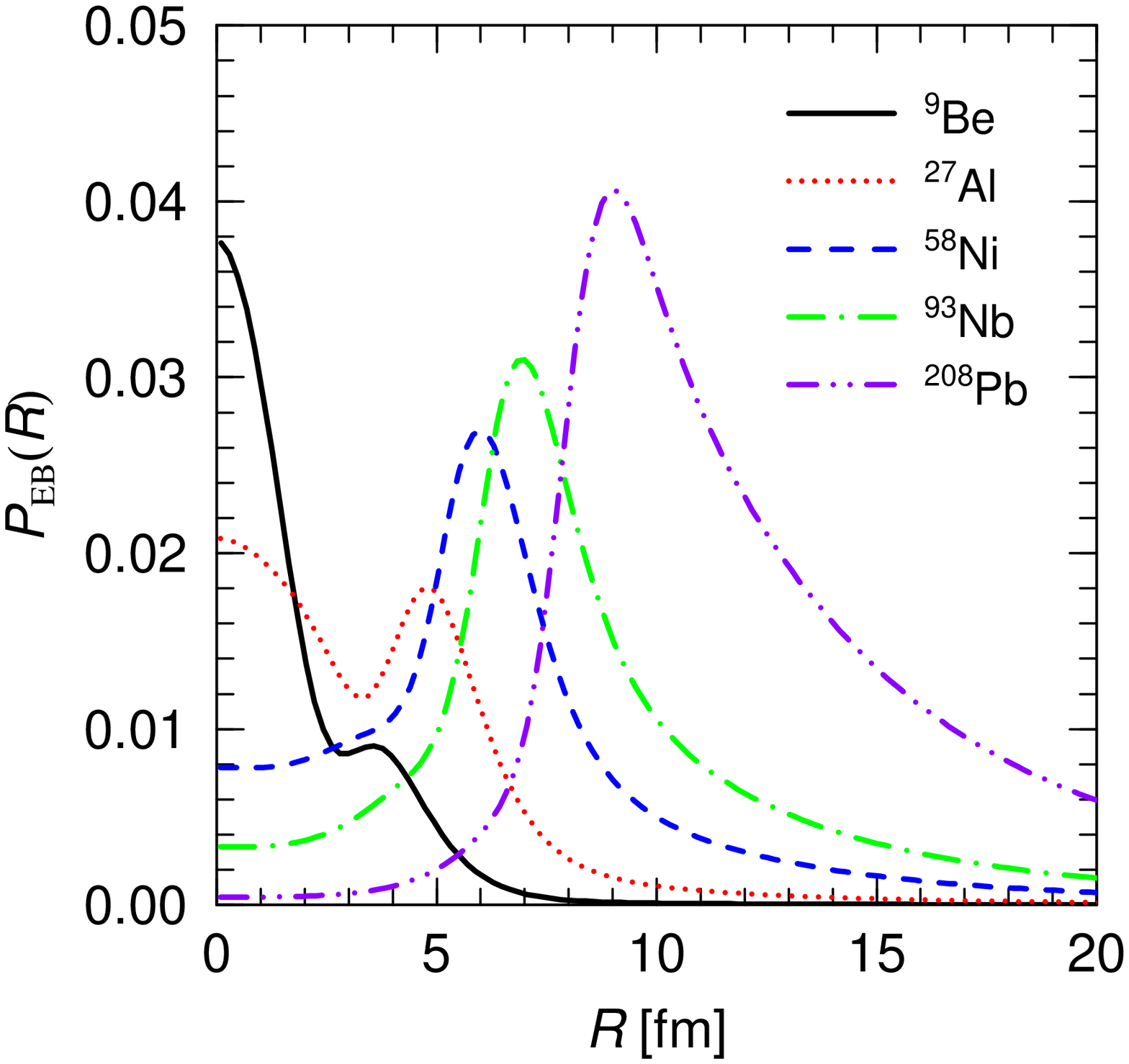}
 \caption{(Color online)
 Transition probabilities $P_{\rm R}$ and $P_{\rm EB}$
 as a function of $R=(J+1/2)/K_0$  for deuteron scattering at
 200~MeV/nucleon.
   }
 \label{Fig-relation}
\end{center}
\end{figure}

The transition probabilities $P_{\rm R}$ and $P_{\rm EB}$ are plotted
in Fig.~\ref{Fig-relation} as a function of
the effective distance $R \equiv (J+1/2)/K_0$
 between the projectile and the target.
For heavier targets, $^{58}$Ni, $^{93}$Nb and $^{208}$Pb,
$P_{\rm R}$ behaves as a logistic function and hence the $R$ dependence is
close to a step function. This indicates that
the reaction cross section can be approximately described
by the black-sphere model~\cite{Kohama}.
Therefore, $\sigma_{\rm R}$ can be expressed by the area of a disk
\bea
\sigma_{\rm R}=\pi R_{\rm R}^2 .
\label{R-reaction}
\eea
with effective radius $R_{\rm R}$.
Meanwhile, the elastic-breakup reaction is peripheral, since
$P_{\rm EB}$ has a single peak at a finite value of $R$.
An effective radius $R_{\rm EB}$ of $\sigma_{\rm EB}$ can be defined by
the peak of $P_{\rm EB}$.
As expected, $P_{\rm R}$ changes rapidly at $R=R_{\rm EB}$.
This indicates that
\bea
R_{\rm R}=R_{\rm EB} .
\label{R-EB-formula}
\eea
For lighter targets such as $^{9}$Be and $^{27}$Al,
$P_{\rm EB}$ has two peaks; the first peak is located at $R=0$ and
the second at finite value of $R$.
However, the second peak is more significant than the first peak
in $\sigma_{\rm R}$ because of the
weight factor of $2J+1$ in Eq.~\eqref{Xsec-R}.
We thus define $R_{\rm EB}$ by the second peak.

\begin{figure}[htbp]
\begin{center}
 \includegraphics[width=0.4\textwidth,clip]{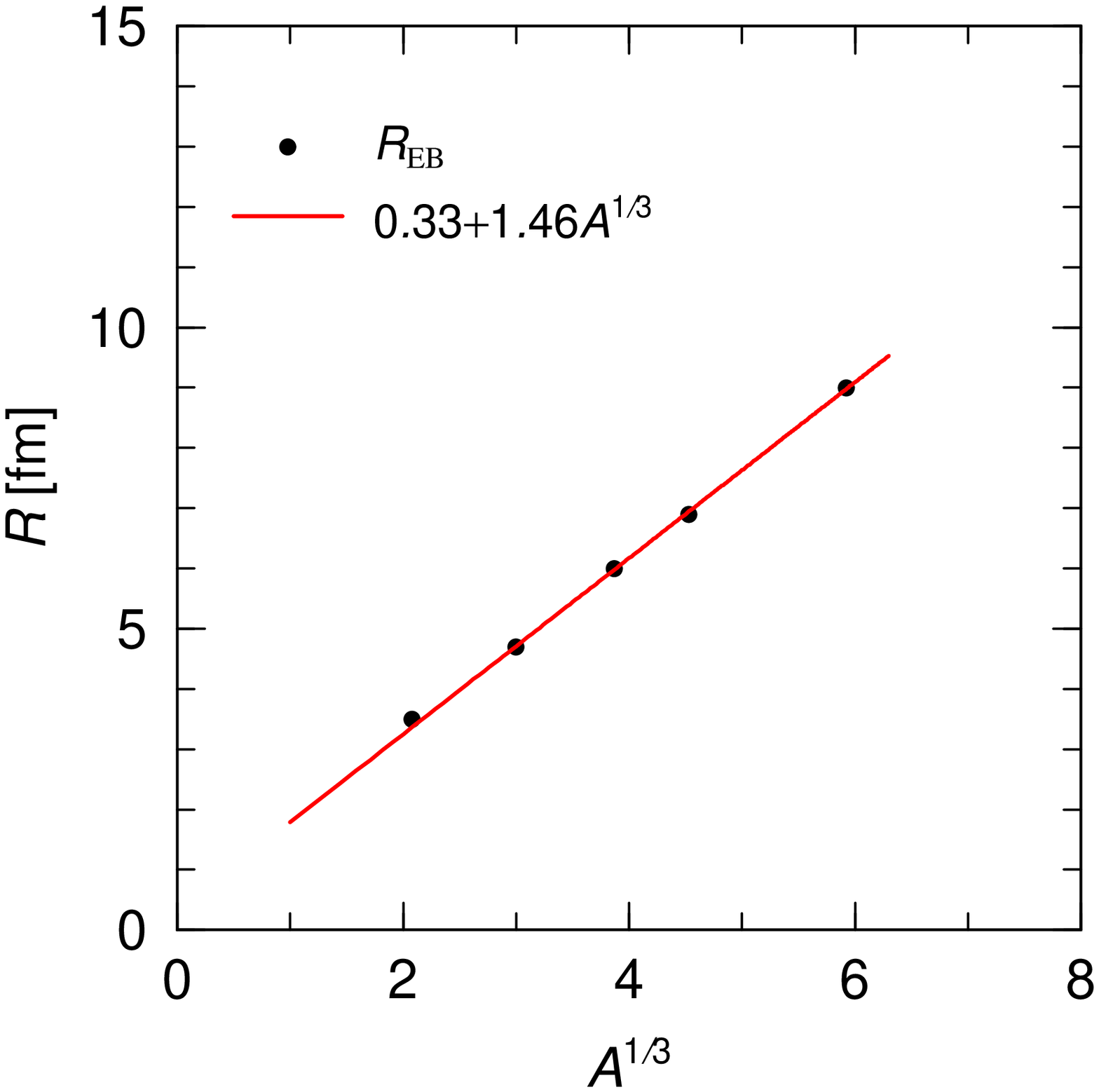}
 \includegraphics[width=0.4\textwidth,clip]{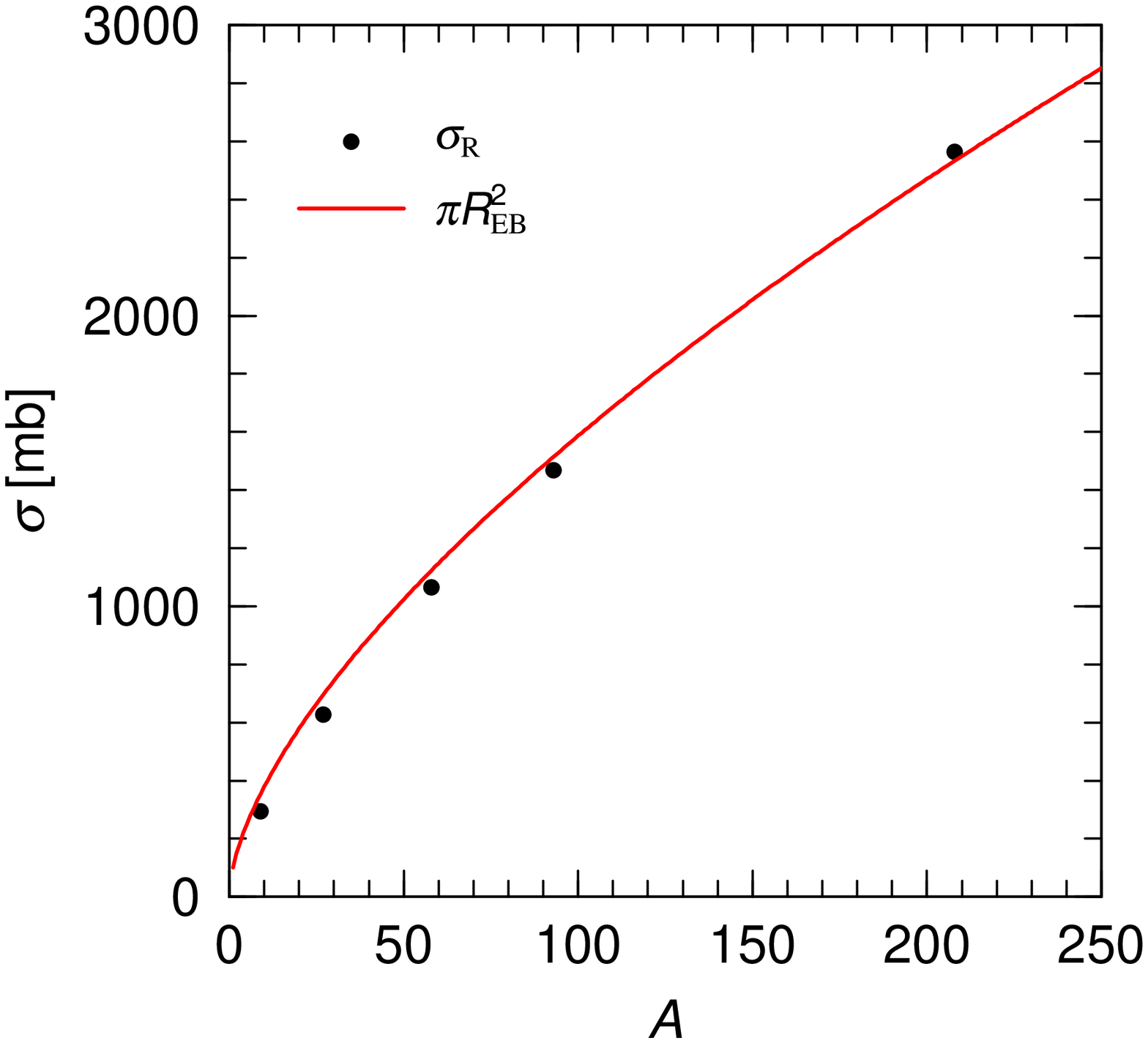}
 \caption{(Color online)
 $A$ dependence of (a) $R_{\rm EB}$ and (b) $\sigma_{\rm R}$.
     }
 \label{Fig-formula}
\end{center}
\end{figure}

Figure~\ref{Fig-formula}(a) shows $R_{\rm EB}$ as a function of $A^{1/3}$,
where $A$ is the target mass number.
Since the elastic-breakup reaction is peripheral, $R_{\rm EB}$ is expected
to depend on $A^{1/3}$. Actually, $A$-dependence of $R_{\rm EB}$ is
well fitted by a straight line (the solid line)
\bea
R_{\rm EB}=0.33+1.46 A^{1/3} .
\label{EB-line}
\eea
The fitting is made only for heavier targets of
$^{58}$Ni, $^{93}$Nb, and $^{208}$Pb, where $P_{\rm EB}(J)$ has a
single peak.
Figure~\ref{Fig-formula}(b) shows $A$-dependence of $\sigma_{\rm R}$,
where the solid curve is obtained from Eq.~\eqref{R-reaction} with
Eqs.~\eqref{R-EB-formula} and \eqref{EB-line}, while
closed circles stand for the results of CDCC.
Both results agree with each other.
Thus, the formula \eqref{R-EB-formula} is well satisfied.

\subsection{$A$-dependence of integrated cross sections}
\label{sec:A-X-sec}

$A$-dependence of integrated cross sections is discussed
for deuteron induced reactions at 200~MeV/nucleon.
Integrated cross sections calculated by CDCC and ERT are tabulated in
Table~\ref{tbl-X-sec}.

\begin{table}[htbp]
\begin{center}
\begin{tabular}{c|ccc|cc|cc|c}
 \hline \hline
 Nuclide & $\sigma_{\rm R}$ & $\sigma_{\rm EB}$ & $\sigma_{\rm TF}$ &
 $\sigma_{n:{\rm STR}}$ & $\sigma_{-n}$ & $\sigma_{p:{\rm STR}}$ &
 $\sigma_{-p}$ & $\sigma_{\rm CF}$ \\
 \hline
 $^9$Be     &  295 &  11 &  284 & 111 & 122 & 133 & 144 &   40 \\
 $^{27}$Al  &  628 &  30 &  598 & 211 & 241 & 245 & 275 &  144 \\
 $^{58}$Ni  & 1066 &  69 &  997 & 312 & 381 & 343 & 412 &  342 \\
 $^{93}$Nb  & 1469 & 110 & 1359 & 387 & 497 & 406 & 516 &  566 \\
 $^{208}$Pb & 2565 & 275 & 2290 & 540 & 815 & 489 & 764 & 1261 \\
 \hline \hline
\end{tabular}
\end{center}
\caption{\label{tbl-X-sec}
Integrated cross sections calculated with CDCC and ERT.
The cross sections are shown in units of mb.
}
\end{table}

First, we consider the total-fusion cross section $\sigma_{\rm TF}$.
The cross section is obtained by subtracting the area of the
ring $\sigma_{\rm EB}(J)$ from that of the disk $\sigma_{\rm R}(J)$.
Thus, it can be described also by the area of a disk
\bea
\sigma_{\rm TF}=\pi R_{\rm TF}^2
\label{R-TF}
\eea
with effective radius $R_{\rm TF}$.
Similar definition is possible for
$\sigma_{\rm TF}(p)$ and $\sigma_{\rm TF}(n)$:
\bea
\sigma_{\rm TF}(p)=\pi R_{\rm TF}(p)^2 ,~~~
\sigma_{\rm TF}(n)=\pi R_{\rm TF}(n)^2 .
\label{R-TF-N}
\eea

\begin{figure}[htbp]
\begin{center}
 \includegraphics[width=0.4\textwidth,clip]{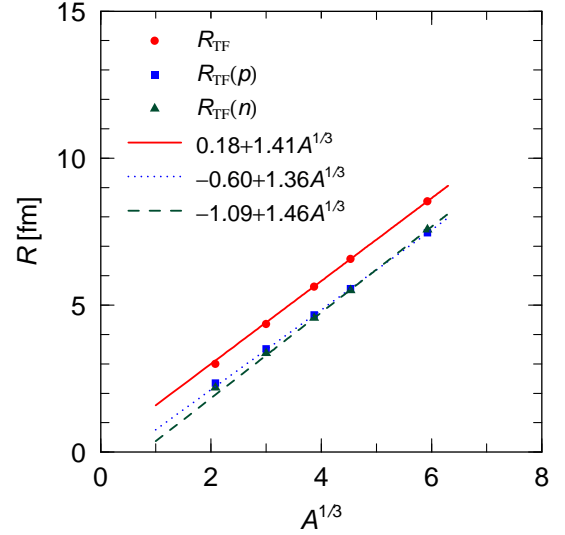}
 \caption{(Color online)
 $A$-dependence of $R_{\rm TF}$, $R_{\rm TF}(p)$ and
 $R_{\rm TF}(n)$.
 The symbols denote the results of CDCC, while the lines stand for the
 results of the straight-line fitting.
   }
 \label{Fig-R-TF}
\end{center}
\end{figure}

Figure~\ref{Fig-R-TF} presents $R_{\rm TF}$, $R_{\rm TF}(p)$ and
$R_{\rm TF}(n)$ as a function of $A^{1/3}$.
Symbols show effective radii evaluated from the CDCC total-fusion
cross sections
in Table~\ref{tbl-X-sec}. They can be fitted by
straight lines
\bea
R_{\rm TF}&=&0.18+1.41 A^{1/3} , \\
\label{TF-line}
R_{\rm TF}(p)&=&-0.60+1.36 A^{1/3} , \\
\label{TF-p-line}
R_{\rm TF}(n)&=&-1.09+1.46 A^{1/3} .
\label{TF-n-line}
\eea
This fitting is made only for heavier targets,
$^{58}$Ni, $^{93}$Nb, and $^{208}$Pb, where $P_{\rm R}(J)$ has a logistic shape,
but the fitting is still good for lighter targets of $^{9}$Be and $^{27}$Al.

The neutron stripping cross section is obtained from
$R_{\rm TF}$ and $R_{\rm TF}(p)$ as
\bea
\sigma_{n:{\rm STR}} = \pi \big[ R_{\rm TF}^2-R_{\rm TF}(p)^2 \big]
                   = 2 \pi D_{n:{\rm STR}} R_{n:{\rm STR}}
\label{R-n:STR}
\eea
with
\bea
D_{n:{\rm STR}} &=& R_{\rm TF}-R_{\rm TF}(p),
\\
R_{n:{\rm STR}} &=& \big[ R_{\rm TF}+ R_{\rm TF}(p) \big]/2 .
\eea
Thus, the neutron stripping reaction occurs on a ring of effective
radius $R_{n:{\rm STR}}$ and effective width $D_{n:{\rm STR}}$.
The effective width $D_{n:{\rm STR}}$ has small $A$-dependence because of the
cancellation between $R_{\rm TF}$ and $R_{\rm TF}(p)$.
Similar discussion can be made for the proton stripping cross section:
\bea
\sigma_{p:{\rm STR}} = 2 \pi D_{p:{\rm STR}} R_{p:{\rm STR}}
\label{R-p:STR}
\eea
with
\bea
D_{p:{\rm STR}} &=& R_{\rm TF}-R_{\rm TF}(n),
\\
R_{p:{\rm STR}} &=& \big[ R_{\rm TF}+ R_{\rm TF}(n) \big]/2 .
\eea

\begin{figure}[htbp]
\begin{center}
 \includegraphics[width=0.4\textwidth,clip]{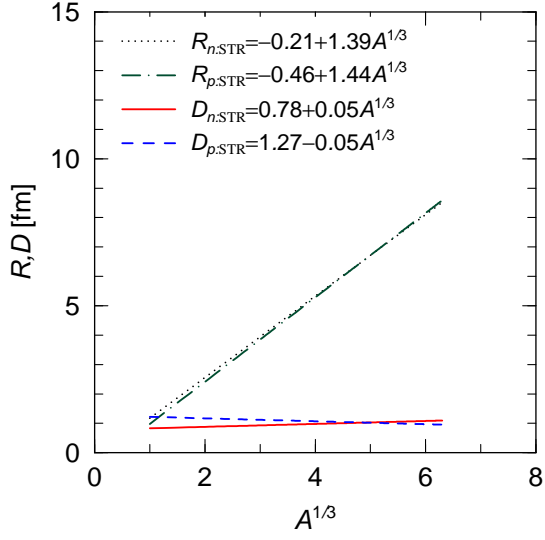}
 \caption{(Color online)
 $A$ dependence of $R_{n:{\rm STR}}$, $R_{p:{\rm STR}}$, 
 $D_{n:{\rm STR}}$, and $D_{p:{\rm STR}}$.
     }
 \label{Fig-STR-DR}
\end{center}
\end{figure}

The effective radii $R_{n:{\rm STR}}$ and $R_{p:{\rm STR}}$, and
the effective widths $D_{n:{\rm STR}}$ and $D_{p:{\rm STR}}$ are simply obtained
from $R_{\rm TF}$, $R_{\rm TF}(p)$, and $R_{\rm TF}(n)$:
\bea
R_{n:{\rm STR}}&=&-0.21+1.39A^{1/3},
\label{STR-RD-1} \\
R_{p:{\rm STR}}&=&-0.46+1.44A^{1/3},
\label{STR-RD-2} \\
D_{n:{\rm STR}}&=&0.78+0.05A^{1/3},
\label{STR-RD-3} \\
D_{p:{\rm STR}}&=&1.27-0.05A^{1/3}.
\label{STR-RD-4}
\eea
These are shown in Fig.~\ref{Fig-STR-DR} as a function
of $A^{1/3}$. As expected, $D_{n:{\rm STR}}$ and $D_{p:{\rm STR}}$ have weak
$A$-dependence. The values are about 1~fm corresponding to
the diffuseness of the target density.
Meanwhile, $R_{n:{\rm STR}}$ and $R_{p:{\rm STR}}$ has
almost the same $A$-dependence as $R_{\rm R}$.

\begin{figure}[htbp]
\begin{center}
 \includegraphics[width=0.4\textwidth,clip]{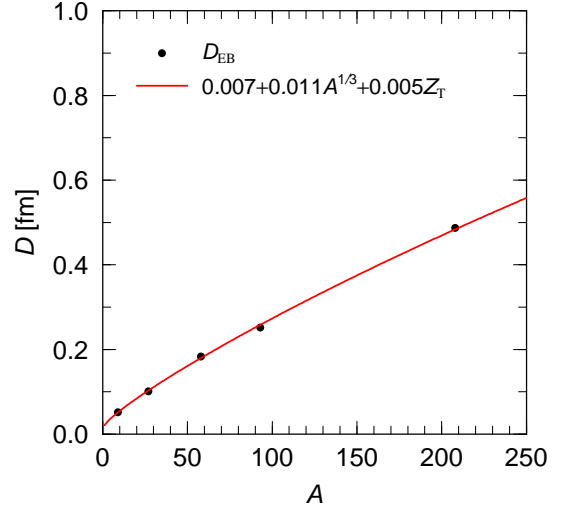}
 \caption{(Color online)
 $A$ dependence of $D_{\rm EB}$.
     }
 \label{Fig-DEB}
\end{center}
\end{figure}

The elastic-breakup reaction is peripheral as
the stripping reactions. It is thus natural to assume that
it occurs on a ring with effective radius $R_{\rm EB}$ and effective
width $D_{\rm EB}$:
\bea
\sigma_{\rm EB}=2 \pi D_{\rm EB} R_{\rm EB} .
\label{R-EB}
\eea
The effective width $D_{\rm EB}$ may be parameterized by
\bea
D_{\rm EB}=a+b A^{1/3}+c Z_{\rm T}
\label{D-EB}
\eea
with parameters $a$, $b$, and $c$.
Note that the role of Coulomb breakup, which is essential
for $\sigma_{\rm EB}$, is described by the last term $c Z_{\rm T}$,
where $Z_{\rm T}$ is the proton number of target.
We use the relation between $Z_{\rm T}$ and $A$ 
for nuclei on stability line:
\bea
Z_{\rm T}=\frac{A}{2+(a_{\rm C}/a_{\rm A})A^{2/3}}
       =\frac{A}{2+0.015A^{2/3}}
\eea
obtained from the Bethe-Weis\"{a}cker mass formula
\cite{MassFormula} neglecting pairing energy term,
where $a_{\rm C}=0.697$ MeV and $a_{\rm A}=46.58$ MeV are
coefficients of the Coulomb and asymmetry energy terms, respectively.
The parameterization \eqref{D-EB} indeed works well
as shown in Fig.~\ref{Fig-DEB}.
The circles denote $D_{\rm EB}$ evaluated from the CDCC
elastic-breakup cross section with Eq.~\eqref{R-EB}.
$A$-dependence of the CDCC results is well simulated by
the solid line
\bea
D_{\rm EB}=0.007 + 0.011 A^{1/3}+ 0.005 Z_{\rm T} .
\label{DEB-line}
\eea
The $D_{\rm EB}$ is found to be
smaller than $D_{n:{\rm STR}}$ and $D_{p:{\rm STR}}$.

\begin{figure}[htbp]
\begin{center}
 \includegraphics[width=0.4\textwidth,clip]{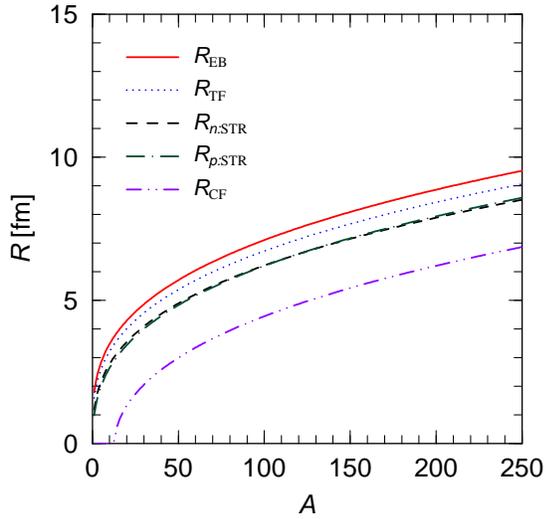}
 \caption{(Color online)
 $A$ dependence of $R_{\rm EB}$, $R_{\rm TF}$,
 $R_{n:{\rm STR}}$, $R_{p:{\rm STR}}$, and $R_{\rm CF}$.
      }
 \label{Fig-Radii}
\end{center}
\end{figure}

Other integrated cross sections can be
obtained by the combination of
$\sigma_{\rm EB}$, $\sigma_{n:{\rm STR}}$, $\sigma_{p:{\rm STR}}$, and
$\sigma_{\rm TF}$:
\bea
\sigma_{-n}&=&2 \pi D_{\rm EB} R_{\rm EB}+2\pi D_{n:{\rm STR}} R_{n:{\rm STR}}, \label{-n-XSEC} \\
\sigma_{-p}&=&2 \pi D_{\rm EB} R_{\rm EB}+2\pi D_{p:{\rm STR}} R_{p:{\rm STR}}, \label{-p-XSEC} \\
\sigma_{\rm IF}&=&2\pi D_{n:{\rm STR}} R_{n:{\rm STR}}
+2\pi D_{p:{\rm STR}} R_{p:{\rm STR}},
\label{IF-XSEC} \\
\sigma_{\rm CF}&=& \pi R_{\rm TF}^2 - \sigma_{\rm IF}.
\label{CF-XSEC}
\eea
One can thus define effective radii $R_{\rm CF}$
of the complete fusion cross sections by
\bea
\sigma_{\rm CF} \equiv \pi R_{\rm CF}^2
\eea
and can evaluate the value of $R_{\rm CF}$ from Eq.~\eqref{CF-XSEC}.
$A$-dependence of
$R_{\rm TF}$, $R_{n:{\rm STR}}$, $R_{p:{\rm STR}}$, $R_{\rm EB}$, and $R_{\rm CF}$
is summarized in Fig.~\ref{Fig-Radii}.
The order of the effective radii is
$ R_{\rm CF}< R_{p:{\rm STR}} \approx R_{n:{\rm STR}}
< R_{\rm TF} < R_{\rm R}=R_{\rm EB}$, independently of $A$.
Among these reactions, the elastic-breakup reaction is most peripheral, and
it occurs at $R_{\rm EB}-D_{\rm EB}/2 \la R \la R_{\rm EB}+D_{\rm EB}/2$.
The incomplete fusion reactions take place at
$R_{n:{\rm STR}}-D_{n:{\rm STR}}/2 \la R \la R_{n:{\rm STR}}+D_{n:{\rm STR}}/2$.
At $R \la R_{\rm CR}$, only the complete fusion reaction occurs.

\begin{figure}[htbp]
\begin{center}
 \includegraphics[width=0.4\textwidth,clip]{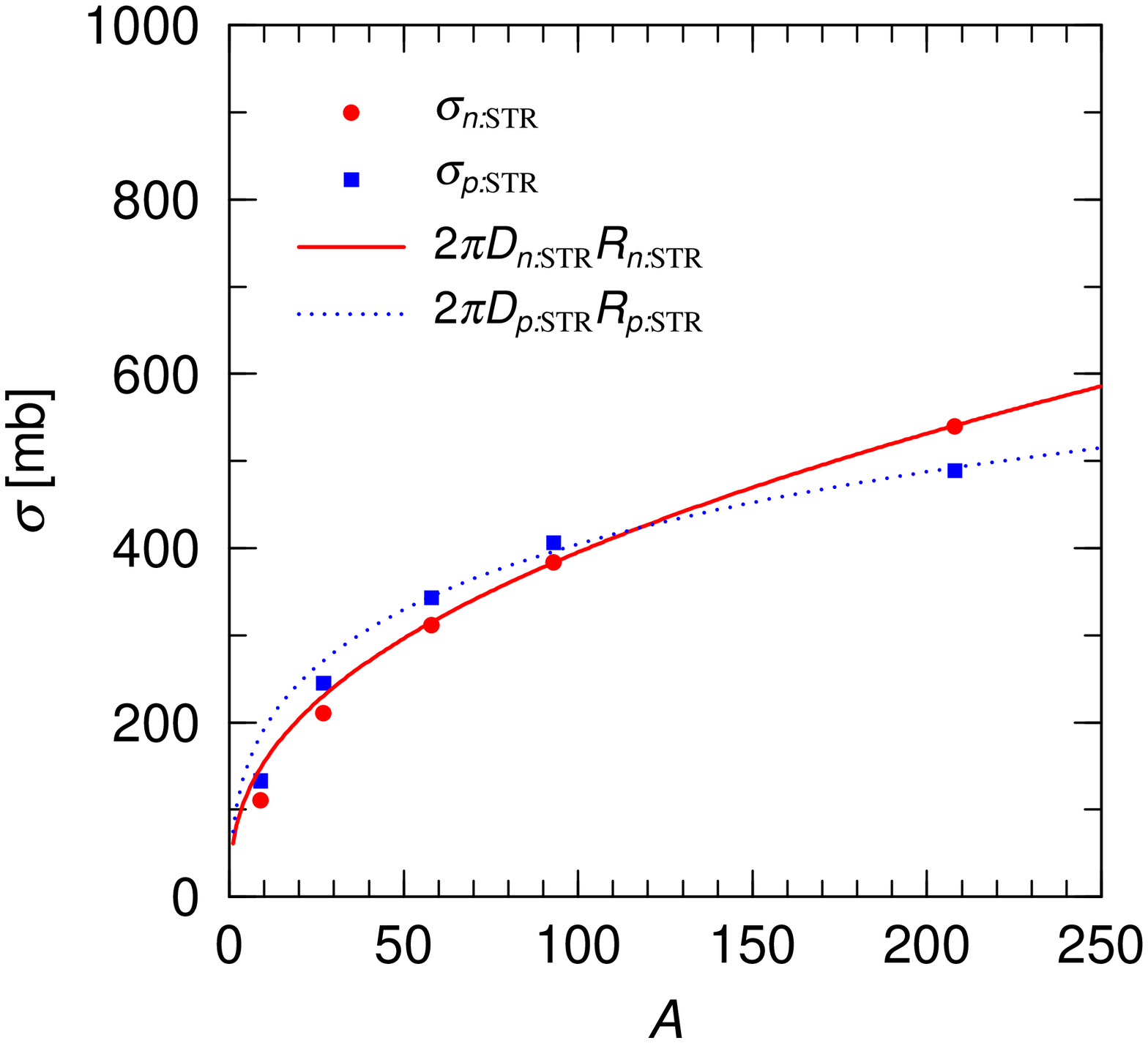}
 \includegraphics[width=0.4\textwidth,clip]{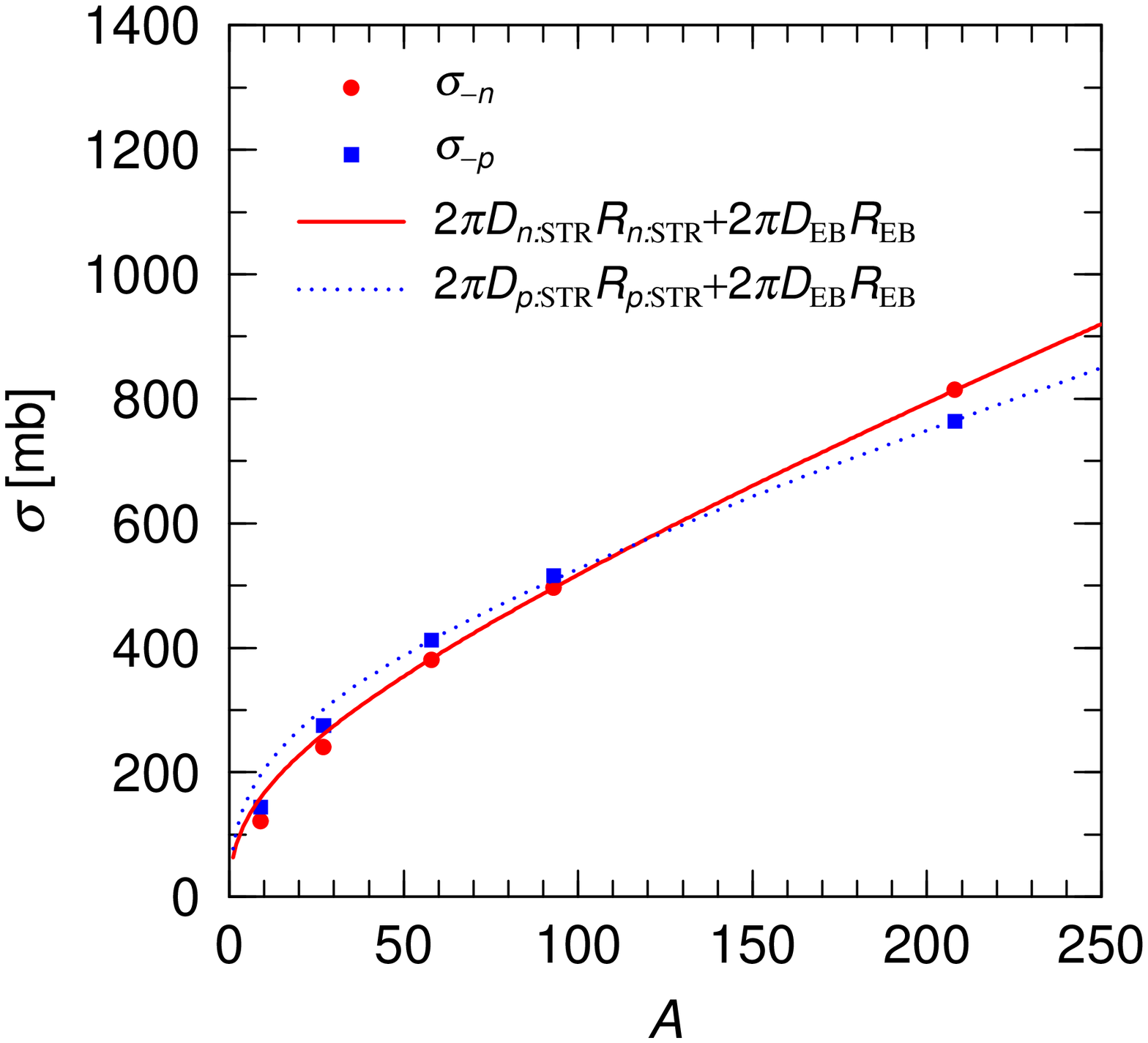}
 \caption{(Color online)
 $A$-dependence of the stripping and removal cross sections.
  $\sigma_{n:{\rm STR}}$ and $\sigma_{p:{\rm STR}}$ are plotted
  in the upper panel,
  while $\sigma_{-n}$ and $\sigma_{-p}$ are shown in the lower panel.
  Symbols stand for results of CDCC and ERT. Lines denote results of
  the parameterization \eqref{R-n:STR}-\eqref{CF-XSEC}. }
 \label{Fig-STR-XSEC}
\end{center}
\end{figure}

The cross sections $\sigma_{n:{\rm STR}}$, $\sigma_{p:{\rm STR}}$,
$\sigma_{-n}$, and $\sigma_{-p}$ are plotted
as a function of $A$ in Fig.~\ref{Fig-STR-XSEC}.
The cross section $\sigma_{n:{\rm STR}}$ ($\sigma_{-n}$)
has similar $A$-dependence to $\sigma_{p:{\rm STR}}$ ($\sigma_{-p}$).
The removal cross sections have stronger $A$-dependence than
the stripping cross sections, since the former include
the elastic-breakup cross section.

\begin{figure}[htbp]
\begin{center}
 \includegraphics[width=0.4\textwidth,clip]{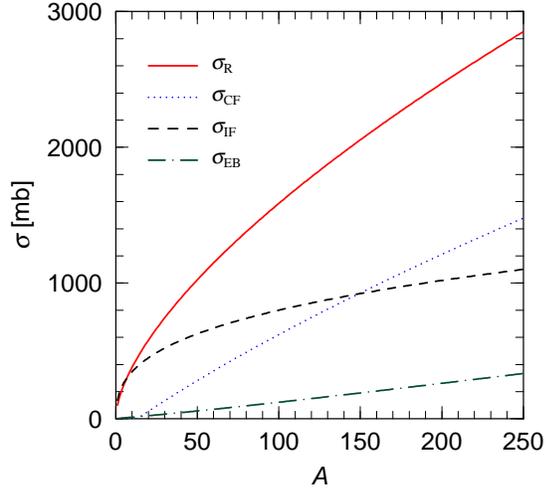}
 \caption{(Color online)
 $A$ dependence of integrated cross sections.
      }
 \label{Fig-XSEC-ALL}
\end{center}
\end{figure}

$A$-dependence of $\sigma_{\rm R}$, $\sigma_{\rm TF}$, $\sigma_{\rm IF}$, and
$\sigma_{\rm CF}$ is summarized
in Fig.~\ref{Fig-XSEC-ALL}.
For $A \la 150$, $\sigma_{\rm IF}$ is larger than $\sigma_{\rm CF}$ and
becomes the largest component of $\sigma_{\rm R}$, while
for $A \ga 150$, $\sigma_{\rm CF}$ becomes the largest.

\subsection{Accuracy of the Glauber model for integrated cross sections}
\label{sec:Glauber}

The accuracy of the Glauber model is investigated
for deuteron induced reactions at 200~MeV/nucleon.
For this purpose, we define the relative error
\bea
\delta_{X}=[X({\rm CDCC})-X({\rm GL})]/X({\rm CDCC}) ,
\eea
where $X({\rm CDCC})$ and $X({\rm GL})$ are integrated cross sections
calculated with CDCC and the Glauber model, respectively.
In the Glauber-model calculation,
the eikonal and adiabatic approximations
are made and the Coulomb interaction is set to zero.

\begin{figure}[htbp]
\begin{center}
 \includegraphics[width=0.4\textwidth,clip]{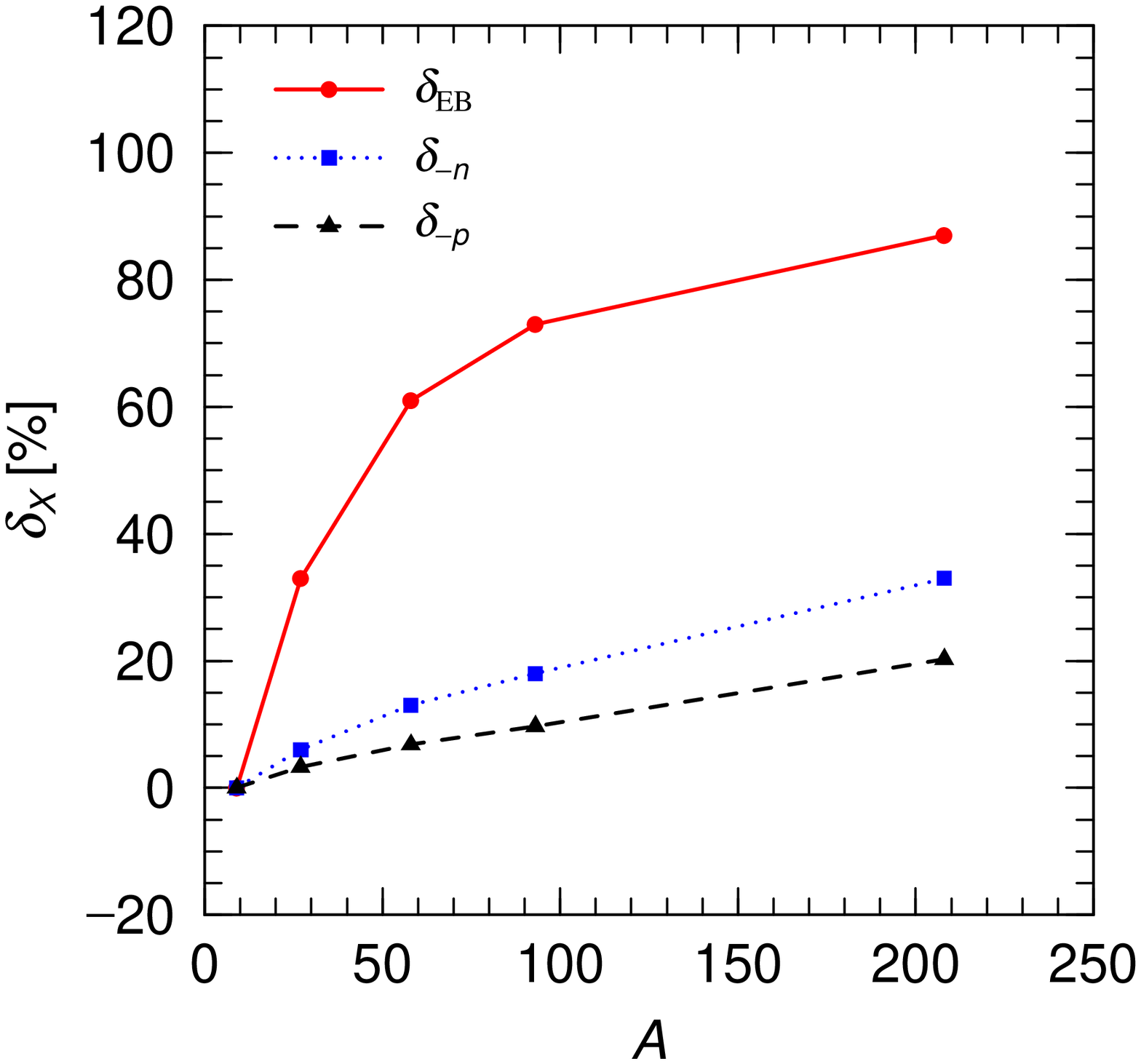}
 \includegraphics[width=0.4\textwidth,clip]{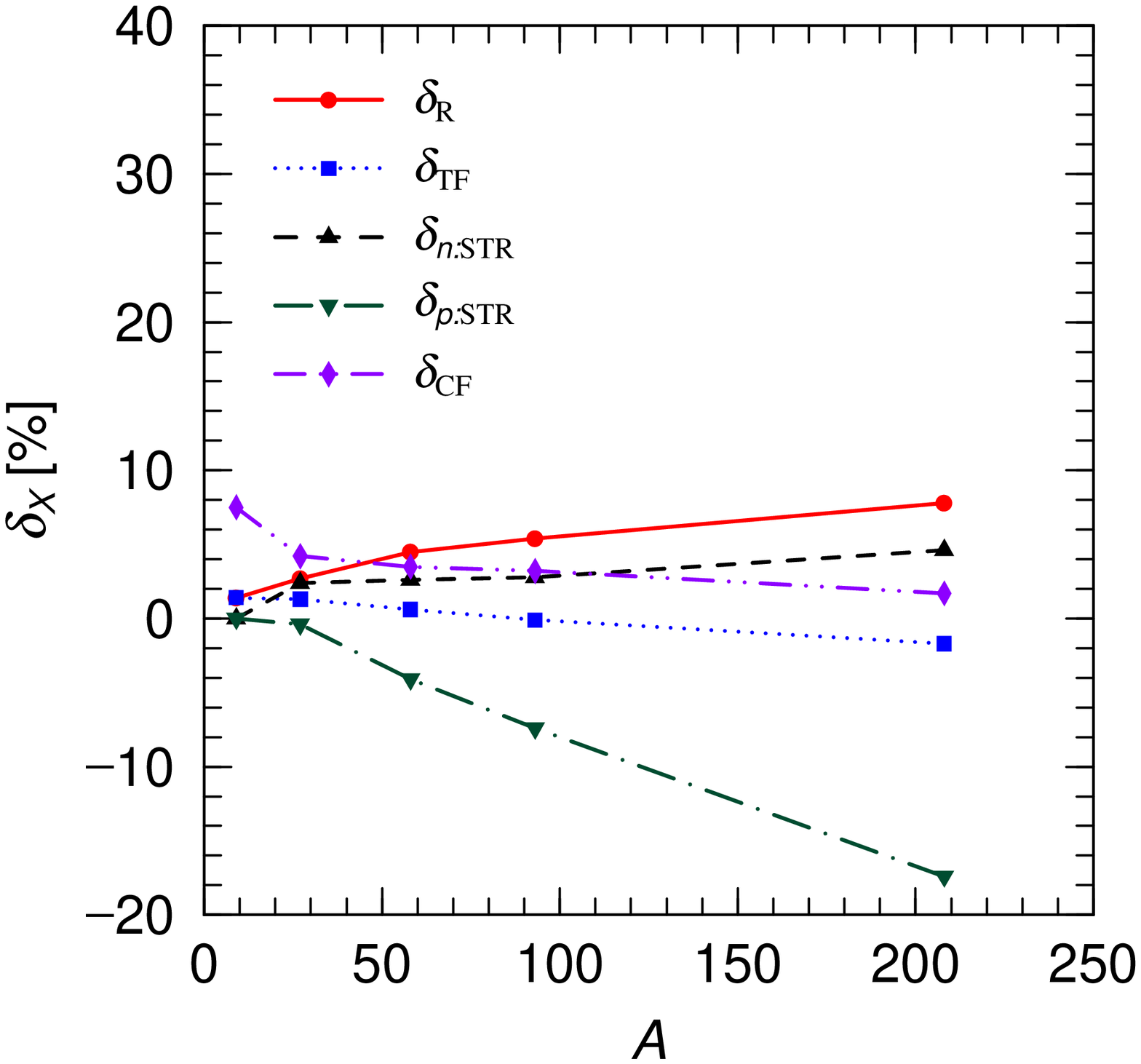}
 \caption{(Color online)
 Accuracy of the Glauber model for integrated cross sections.
        }
 \label{Fig-GL-error}
\end{center}
\end{figure}

Figure~\ref{Fig-GL-error} shows $\delta_{X}$ as a function of $A$
for $\sigma_{\rm EB}$, $\sigma_{-n}$,
and $\sigma_{-p}$ in the upper panel and
for  $\sigma_{\rm R}$, $\sigma_{\rm TF}$,
$\sigma_{n:{\rm STR}}$,  $\sigma_{p:{\rm STR}}$, and $\sigma_{\rm CF}$
in the lower panel.
For light targets, say $^{9}$Be, the error is less than 2\% for
all integrated cross sections except $\sigma_{\rm CF}$. The error is 8\%
for $\sigma_{\rm CF}$, but $\sigma_{\rm CF}$ itself is small there.
Thus, the Glauber model is good at small $A$, as expected.

For heavier targets, say $^{208}$Pb, where the Coulomb breakup is essential,
the error is 80\% for $\sigma_{\rm EB}$, 20\%
for $\sigma_{-n}$ and $\sigma_{-p}$, and $-20$\% for $\sigma_{p:{\rm STR}}$.
The error is slightly smaller for $\sigma_{-p}$
than for $\sigma_{-n}$. However,
this is just a result of the cancellation in $\sigma_{-p}$
between the positive error for $\sigma_{\rm EB}$ and
the negative error for $\sigma_{p:{\rm STR}}$.
The error is less than 6\% for $\sigma_{n:{\rm STR}}$, $\sigma_{\rm R}$,
$\sigma_{\rm TF}$, and $\sigma_{\rm CF}$. Thus, the Glauber model
is not good for $\sigma_{\rm EB}$, $\sigma_{-n}$, and $\sigma_{-p}$.

\section{Summary}
\label{Summary}

The continuum-discretized coupled-channels method (CDCC)
and the eikonal reaction theory (ERT)
are applied to $d+^{58}$Ni elastic scattering at
200~MeV/nucleon and inclusive $^{7}$Li$(d,n)$ reaction at 40~MeV.
For $\sigma_{\rm R}$ of the $d+^{58}$Ni scattering,
the CDCC result is consistent with the experimental data.
The spin-orbit interactions of the proton and neutron optical potentials
yield a significant effect on the differential elastic-scattering
cross section, but not on $\sigma_{\rm R}$.
For $\sigma_{p:{\rm STR}}$ of the $d+^{7}$Li scattering,
the ERT result is consistent with the experimental data.

$A$-dependence of several types of integrated
cross sections is systematically investigated with CDCC and ERT for the
deuteron induced reactions at 200~MeV/nucleon that corresponds to
typical RIBF and GSI beam energies.
The $A$-dependence is clearly explained with simple formulae as follows.
A black-sphere type reaction such as the
reaction, total-fusion, and
complete-fusion processes occurs on a disk with the area of
$\pi R_{\rm X}^2$ with
effective radius $R_{\rm X}$ that is well parameterized by
$a_{\rm X}+b_{\rm X}A^{1/3}$. A peripheral reaction such as the
elastic-breakup,
nucleon-stripping, and nucleon-removal processes takes place
on a ring $2\pi D_{\rm X} R_{\rm X}$
with effective radius and width, $R_{\rm X}$ and $D_{\rm X}$, and
$A$-dependence of $R_{\rm X}$ is well parameterized by
$a_{\rm X}+b_{\rm X}A^{1/3}$.
For neutron- and proton-stripping reactions as the incomplete-fusion reaction,
the effective widths, $D_{n:{\rm STR}}$ and $D_{p:{\rm STR}}$, are
about 1~fm independently of $A$.
The effective radii, $R_{n:{\rm STR}}$ and $R_{p:{\rm STR}}$, are smaller than
effective radius $R_{\rm R}$
for the reaction cross section
by about 1~fm independently of $A$, while
effective radius $R_{\rm CF}$ for the complete-fusion cross section is
smaller than $R_{\rm R}$ by about 2.5~fm independently of $A$.
Thus, $A$-dependences of $R_{\rm R}$, $R_{n:{\rm STR}}$, $R_{p:{\rm STR}}$,
and $R_{\rm CF}$ are
simple and similar to each other.
Thus, if $\sigma_{\rm R}$, $\sigma_{\rm CF}$, $\sigma_{n:{\rm STR}}$, and
$\sigma_{p:{\rm STR}}$ are determined experimentally just for two targets,
one can estimate these cross sections for any target.
It is of interest as a future work to see whether
this property is held for other incident energies and projectiles.

The total-fusion cross section $\sigma_{\rm TF}$ is obtained
from measurable cross sections, $\sigma_{\rm R}$ and $\sigma_{\rm EB}$,
by $\sigma_{\rm TF}=\sigma_{\rm R}-\sigma_{\rm EB}$.
Similarly, neutron- and proton-stripping cross sections are
determined from measurable neutron- and proton-removal cross sections
by $\sigma_{n:{\rm STR}}=\sigma_{-n} - \sigma_{\rm EB}$ and
$\sigma_{p:{\rm STR}}=\sigma_{-p} - \sigma_{\rm EB}$.
It is thus important to determine $A$-dependence of $\sigma_{\rm EB}$.
However, the $A$-dependence is known to be complicated~\cite{Ogata:2008ke,Hussein2},
since it depends on not only $A$ but also
the target proton number $Z_{\rm T}$. This problem can be solved
by the formula $\sigma_{\rm EB}=2\pi D_{\rm EB} R_{\rm EB}$.
Effective radius $R_{\rm EB}$ agrees with
$R_{\rm R}=a_{\rm R}+b_{\rm R}A^{1/3}$ with high accuracy, and
effective width
$D_{\rm EB}$ is well parameterized
by $a_{\rm EB}+b_{\rm EB}A^{1/3}+c_{\rm EB}Z_{\rm T}$.
Thus, $A$-dependence of $\sigma_{\rm EB}$ is determined, if
$\sigma_{\rm EB}$ is measured for three targets and
$\sigma_{\rm R}$ is measured for two targets.

Accuracy of the Glauber model is also tested for the deuteron scattering at
200~MeV/nucleon.
The accuracy for integrated cross sections is summarized as follows.
The Glauber model is good for light targets, if
the interactions between projectile and target are clearly determined.
For heavy targets, however, the model is not good for
the elastic-breakup, the nucleon-removal, and the proton-stripping cross
sections, because of the strong Coulomb field, while it is fairly good
for the other cross sections.
It is quite interesting as a future work that similar systematic analyses
will be made for heavier projectiles such as Ne and Ca isotopes with larger
proton numbers.

\section*{Acknowledgements}
We are grateful to Y.~Suzuki for useful discussions.
We also acknowledge K.~Hagino for valuable suggestions.
S.H. would like to thank A.~Kohama for helpful comments and discussions.



\begin{thebibliography}{00}

\bibitem{Gade}
A.~Gade, \textit{et al.},
Phys.\ Rev.\ C {\bf 77}, 044306 (2008).

\bibitem{Matsui}
H.~Matsui, in
{\it Proceedings of the 23rd Symposium on Fusion Technology},
Venice, Italy, 20-24 Sept. (2004).

\bibitem{Glauber}
R.J.~Glauber, {\it in Lectures in Theoretical Physics}
(Interscience, New York, 1959), Vol. 1, p.315.

\bibitem{Yahiro-Glauber}
M.~Yahiro, K.~Minomo, K.~Ogata, and M.~Kawai,
Prog.\ Theor.\ Phys.\ {\bf 120}, 767(2008).

\bibitem{Capel-08}
P. Capel, D. Baye, and Y. Suzuki,
Phys.\ Rev.\ C {\bf 78}, 054602 (2008).

\bibitem{Tostevin}
J.S. Al-Khalili and J.A. Tostevin,
Phys. Rev. Lett. {\bf 76}, 3903 (1996);
J.S. Al-Khalili, J.A. Tostevin, and I.J. Thompson,
Phys. Rev. C {\bf 54}, 1843 (1996).

\bibitem{Hencken}
K. Hencken, G. Bertsch, and H. Esbensen,
Phys. Rev. C {\bf 54}, 3043 (1996).

\bibitem{Ogawa01}
Y. Ogawa, T. Kido, K. Yabana, and Y. Suzuki,
Prog. Theor. Phys. Suppl. {\bf 142}, 157 (2001),
and references cited therein.

\bibitem{Ibrahim}
B. Abu-Ibrahim and Y. Suzuki,
Prog. Theor. Phys. {\bf 112}, 1013 (2004);
B. Abu-Ibrahim and Y. Suzuki,
Prog. Theor. Phys. {\bf 114}, 901 (2005).

\bibitem{30Ne}
W. Horiuchi, Y. Suzuki, P. Capel, and D. Baye,
Phys. Rev. C {\bf 81}, 024606 (2010).

\bibitem{Hagiwara}
M.~Hagiwara, T.~Itoga, N.~Kawata, N.~Hirabayashi, T.~Oishi,
T.~Yamauchi, M.~Baba, M.~Sugimoto, and T.~Muroga,
Fusion Sci. Technol. {\bf 48}, 1320 (2005).

\bibitem{Ye}
T.~Ye, Y.~Watanabe, and K.~Ogata,
Phys. Rev. C {\bf 80}, 014604 (2009).

\bibitem{CDCC-review1}
M.~Kamimura, 
M.~Yahiro, Y.~Iseri, Y.~Sakuragi, H.~Kameyama, and
M.~Kawai, \newblock
Prog.\ Theor.\ Phys.\ Suppl.\ {\bf 89}, 1 (1986).

\bibitem{CDCC-review2}
N.~Austern, 
Y.~Iseri, M.~Kamimura, M.~Kawai, G.~Rawitscher, and
M.~Yahiro, \newblock
Phys.\ Rep.\ {\bf 154}, 125 (1987).

\bibitem{CDCC-foundation1}
N.~Austern, M.~Yahiro, and M.~Kawai,
\newblock
Phys.\ Rev.\ Lett. {\bf 63}, 2649 (1989).

\bibitem{CDCC-foundation2}
N.~Austern, 
M.~Kawai, and  M.~Yahiro, \newblock
Phys.\ Rev.\ C {\bf 53}, 314 (1996).

\bibitem{CDCC-foundation3}
A. Deltuva, A.M.~Moro, E.~Cravo, F.M.~Nunes, and A.C.~Fonseca,
Phys.\ Rev.\ C {\bf 76}, 064602 (2007).

\bibitem{Rusek}
K. Rusek and K.W. Kemper,
Phys. Rev. C {\bf 61}, 034608 (2000).

\bibitem{Tostevin2}
J.A. Tostevin, F.M. Nunes, and I.J. Thompson,
Phys. Rev. C {\bf 63}, 024617 (2001).

\bibitem{Davids}
B. Davids, S.M. Austin, D. Bazin, H. Esbensen,
B.M. Sherrill, I.J. Thompson, and J.A. Tostevin,
Phys. Rev. C {\bf 63}, 065806 (2001).

\bibitem{Mortimer}
J. Mortimer, I. J. Thompson, and J. A. Tostevin,
Phys.\ Rev.\ C {\bf 65}, 064619 (2002).

\bibitem{Eikonal-CDCC}
K.~Ogata, M.~Yahiro, Y.~Iseri, T.~Matsumoto, and M.~Kamimura,
Phys.\ Rev.\ C {\bf 68}, 064609 (2003).

\bibitem{Matsumoto3}
T.~Matsumoto, 
E.~Hiyama, K.~Ogata, Y.~Iseri, M.~Kamimura,
S.~Chiba, and M.~Yahiro,
Phys.\ Rev.\ C {\bf 70}, 061601(R) (2004).

\bibitem{Howell}
D.J. Howell, J.A. Tostevin, and J.S. Al-Khalili,
J. Phys. G: Nucl. Part. Phys. {\bf 31}, S1881 (2005).

\bibitem{Rusek2}
K. Rusek, I. Martel, J. G\'{o}mez-Camacho,
A.M. Moro, and R. Raabe,
Phys. Rev. C {\bf 72}, 037603 (2005).

\bibitem{Matsumoto4}
T.~Matsumoto, 
T.~Egami, K.~Ogata, Y.~Iseri, M.~Kamimura, and M.~Yahiro,
Phys.\ Rev.\ C {\bf 73}, 051602(R) (2006).

\bibitem{Moro}
A.M. Moro, K. Rusek, J.M. Arias, J. G\'{o}mez-Camacho,
and M. Rodr\'{i}guez-Gallardo,
Phys. Rev. C {\bf 75}, 064607 (2007).

\bibitem{THO-CDCC}
M.~Rodr\'{i}guez-Gallardo, 
J.~M.~Arias, J.~G\'{o}mez-Camacho,
R.~C.~Johnson, A.~M.~Moro, I.~J.~Thompson, and J.~A.~Tostevin,
\newblock
Phys.\ Rev.\ C {\bf 77}, 064609 (2008).

\bibitem{4body-CDCC-bin}
M.~Rodr\'{i}guez-Gallardo,
J. M. Arias, J. G\'{o}mez-Camacho,
A. M. Moro, I. J. Thompson, and J. A. Tostevin,
Phys.\ Rev.\ C {\bf 80}, 051601(R) (2009).

\bibitem{Matsumoto:2010mi}
T.~Matsumoto, K.~Kato, and M.~Yahiro,
Phys.\ Rev.\  C {\bf 82}, 051602 (2010)
[arXiv:1006.0668 [nucl-th]].

\bibitem{Avrigeanu}
M. Avrigeanu and A.M. Moro,
Phys. Rev. C {\bf 82}, 037601 (2010).

\bibitem{ERT}
M. Yahiro, K. Ogata, and K. Minomo,
arXiv:1103.3976 (2011) [nucl-th].

\bibitem{Hussein}
M. S. Hussein and K. W. McVoy,
Nucl. Phys. {\bf A445}, 124 (1985).

\bibitem{KDpot}
A.J.~Koning and J.P.~Delaroche,
Nucl. Phys. {\bf A713}, 231 (2003).

\bibitem{Ohmura}
T.~Ohmura, B.~Imanishi, M.~Ichimura, and M.~Kawai,
Prog.\ Theor.\ Phys.\ {\bf 43}, 347 (1970).

\bibitem{d58Ni}
N.van Sen \textit{et al.},
Phys. Lett. {\bf B156}, 185 (1985).

\bibitem{Kohama}
A.~Kohama, K.~Iida, and K.~Oyamatsu,
Phys. Rev. C {\bf 72}, 024602 (2005).

\bibitem{MassFormula}
C.F. von Weizs\"{a}cker, Z. Phys. {\bf 96}, 431 (1935);
H.A. Bethe and R. F. Bacher, Rev. Mod. Phys. {\bf 8}, 82 (1936).

\bibitem{Hussein2}
M.S. Hussein, R. Lichtenth\"aler, F.M. Nunes, and I.J. Thompson,
Phys. Lett. {\bf B640}, 91 (2006).

\bibitem{Ogata:2008ke}
K.~Ogata, T.~Matsumoto, Y.~Iseri, and M.~Yahiro,
J.\ Phys.\ Soc.\ Jap.\  {\bf 78}, 084201 (2009)
[arXiv:0804.1186 [nucl-th]].

\end{thebibliography}
\end{document}